\documentclass[12pt]{iopart}
\pdfoutput=1
\usepackage{setstack}
\usepackage{iopams}
\eqnobysec

\usepackage{placeins}
\newcommand{\ra}[1]{\renewcommand{\arraystretch}{#1}}

\usepackage{graphicx}
\usepackage[lofdepth=1, lotdepth]{subfig}
\usepackage{tikz}
\usetikzlibrary{chains}
\usetikzlibrary{automata,arrows,positioning,calc,fit}

\usepackage{textcomp}

\usepackage{pgfplots}
\newenvironment{customlegend}[1][]{%
    \begingroup
    \csname pgfplots@init@cleared@structures\endcsname
    \pgfplotsset{#1}%
}{%
    \csname pgfplots@createlegend\endcsname
    \endgroup
}%

\def\addlegendimage{\csname pgfplots@addlegendimage\endcsname}

\definecolor{bleu}{HTML}{FDA422}
\definecolor{darkvert}{HTML}{078407}
\definecolor{rouge}{HTML}{FFFE2C}
\definecolor{jaunefonce}{HTML}{FDA422}
\definecolor{vert}{HTML}{9ccc3b}
\definecolor{jaune}{HTML}{FFFE2C}
\definecolor{orange}{HTML}{FDA422}
\definecolor{orangefonce}{HTML}{EC762D}
\definecolor{rouge}{HTML}{FC0B1B}
\definecolor{rougefonce}{HTML}{890309}

\pgfplotsset{compat=1.5}

\begin{document}

\title[Network model - Propagation of Hepatitis C]{A network model for the propagation of Hepatitis~C with HIV co-infection}

\author{Arnaud Nucit$^{1,2}$ and Julien Randon-Furling$^3$}

\address{$^1$ D\'epartement Strat\'egie et Op\'erations M\'edicales, Bristol-Myers Squibb, Rueil-Malmaison, France}
\address{$^2$ INSA et Institut de Math\'ematiques de Toulouse (UMR~5219), Toulouse, France}
\address{$^3$ SAMM (EA4543), Universit\'e Paris-1 Panth\'eon-Sorbonne,\\ Centre Pierre Mend\`es-France, 90 rue de Tolbiac, 75013 Paris, France}
\ead{Julien.Randon-Furling@univ-paris1.fr}

\begin{abstract}
We define and examine a model of epidemic propagation for a virus such as Hepatitis~C (with HIV co-infection) on a network of networks, namely the network of French urban areas. One network level is that of the individual interactions inside each urban area. The second level is that of the areas themselves, linked by individuals travelling between these areas and potentially helping the epidemic spread from one city to another. We choose to encode the second level of the network as extra, special nodes in the first level. We observe that such an encoding leads to sensible results in terms of the extent and speed of propagation of an epidemic, depending on its source point.
\end{abstract}

%
\vspace{2pc}

\maketitle
%
%

\section{Introduction}

Modelling the propagation of epidemics is a long-standing endeavour in the medical and mathematical sciences. In the past decade or so, multiple developments in the science of complex systems in general, and networks in particular, have led to new models, along with the possibility to test their predictions on datasets, the size of which keeps increasing (see for instance \cite{meta2,meta} and references therein).

Particularly challenging are the study and modeling of epidemics involving specific, sometimes marginal, segments of the general population. For instance, viruses and diseases that are particularly prevalent among people who inject drugs (PWID). Things become even more involved if one tries to take into account correlations between interrelated epidemics, such as Hepatitis~C virus (HCV) and AIDS (HIV) among PWID populations~\cite{coinf}.

In this paper, we introduce a simple model of propagation for HCV, with potential HIV co-infection, on a network of networks, representing PWID populations in distinct but interconnected urban areas. This model is intended as a first step towards a more comprehensive model, and to serve as a basis for comparison with public health statistical data when it becomes available.

In our model, the PWID population of each urban area is a network of individuals, and the fact that people may travel or even move houses from one city to another makes for a second level of links, between cities themselves. We choose to represent this second level via the introduction of special nodes at the first level, that will represent at the individual level in each city, contacts with individuals from other cities. This means that if cities $A$ and $B$ are linked (in the sense that a number of individuals travel from one of them to the other), then we add a node in the network representing individuals in city $A$. The degree and other characteristics of this special node in city $A$ will depend on the number of individuals from city $B$ travelling to city $A$ and on the size of the infected population in city $B$. This makes for a simple treatment of the links between cities. We also adapt the standard susceptible-infected-recovered (SIR) equations to reflect the specifics of the HCV epidemics with possible HIV co-infection, as well as to take into account the special type of network with which we are working. The construction of this simple model is detailed in Section~2 and~3.

In Section 4, we examine, via numerical simulations, how our model works, first on a toy-country with just three cities and then on the network of France's 100 largest urban areas. We find that simulations run for different values of the parameters lead to sensible stylized facts in terms of the extent and speed of propagation of an HCV epidemics in France.

\newpage

\section{Epidemic models}

\subsection{Initial model of the spread of Hepatitis~C}\label{hcvsection}

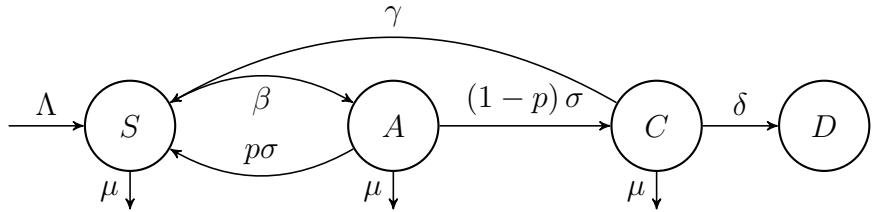
\begin{figure}
\begin{center}
\begin{tikzpicture}[->, >=stealth', auto, semithick, node distance=3.5cm]
\tikzstyle{every state}=[fill=white,draw=black,thick,text=black,scale=1,minimum height = 0.2cm, minimum width = 1.2cm]
\node[state]    (A)  {$S$};
\node[state]    (B)[right of=A]   {$A$};
\node[state]    (C)[right of=B]   {$C$};
\node[state]    (D)[right = 1.0cm of C]   {$D$};
\coordinate[below = 0.5cm of A] (E);
\coordinate[below = 0.5cm of B] (F);
\coordinate[below = 0.5cm of C] (G);
\coordinate[right = 1.0cm of C] (I);
\coordinate[left = 1.0cm of A] (H);
\draw[->] (H)  to   node[above]   {$\Lambda$} (A);
\draw[->] (A)  to   node[left]   {$\mu$} (E);
\draw[->] (B)  to   node[left]   {$\mu$} (F);
\draw[->] (C)  to   node[left]   {$\mu$} (G);
\draw[->] (C)  to   node[above]   {$\delta$} (I);
\path
(A)  edge[ bend left]   node[below]{$\beta$}    (B)
(B)   edge[ left]     node[above]{$\left(1-p\right)\sigma$}     (C)
        edge[ bend left]     node[above]{$p\sigma$}     (A)
(C) edge[ bend right = 30]     node[above]{$\gamma$}     (A);
\end{tikzpicture}
\caption[Compartmental model of HCV's natural history]{\label{markhcv}\textbf{Compartmental model of HCV's natural history}.\\ $S$, susceptible; $A$, acute HCV infection; $C$, chronic HCV infection; $D$, death from HCV. }\end{center}
\end{figure}

Let us consider the compartmental model describing the infection dynamics of Hepatitis~C virus (HCV) displayed in Figure~\ref{markhcv}. This model is based on a susceptible-infected-susceptible (SIS) model which is a derivative of the classic susceptible-infected-recovered (SIR) model introduced in \cite{Kermack}.

We denote by $S$ the number of susceptible individuals, by $A$ the number of individuals in the acute phase of the infection and by $C$ the number of individuals in the chronic phase of the infection. Assuming that the number of individuals in the total population is constant over time, which is not a strong assumption for large populations, we write  $S(t)+A(t)+C(t)=N$. In particular, we assume that the number of people who died from HCV or from some other cause equals the number of people joining the population as new susceptibles.

From now on, we work in terms of proportions rather than absolute numbers. The proportion of susceptible, acute-HCV infected and chronic-HCV infected at time $t$  are denoted by $s(t), a(t)$ and $c(t)$ respectively. Writing $\beta$ for the transmission rate, $\gamma$ for the recovery rate, $\delta$ for the death rate, $\mu$ for the natural mortality rate, $\Lambda$ for the joining rate, $1/\sigma$ for the duration of the acute phase and $p$ for the spontaneous clearance, i.e. the probability that a given individual clears the infection, one can write the following set of differential equations:

\begin{eqnarray}\label{comparthcv}
\eqalign{
\frac{  \mathrm{d}s(t)}{  \mathrm{d}t} = \Lambda -\beta s(t) \big( a(t)+c(t) \big)+ p \sigma a(t) + \gamma c(t) - \mu s(t), \\
\frac{  \mathrm{d}a(t)}{  \mathrm{d}t} = \beta s(t) \big( a(t)+c(t)\big)  - (\sigma + \mu)a(t),\\
\frac{  \mathrm{d}c(t)}{  \mathrm{d}t} = (1-p)\sigma a(t) -(\gamma +\delta+\mu) c(t).} 
\end{eqnarray}
\\

Since HCV is an asymptomatic disease, very limited individuals are diagnosed during the acute phase of the infection. Thus we assume that only individuals in the chronic phase have access to a treatment, and only people in the chronic phase can recover.

Note that the transmission rate $\beta$ may be written as $\beta=\omega \times \pi$, where $\omega$ is the contact rate (the number of contacts per unit time) and $\pi$ the transmission risk (the risk of infection for a given contact). Moreover, as not every patient is actually treated, $\gamma$ may be written as $\gamma= \nu \times q$ where $\nu$ is the recovery rate per patient and $q$ the proportion of treated patients.

The model defined in the set of equations (\ref{comparthcv}), also called mass action model, offers a simple and convenient manner of describing the dynamics of an epidemic in a large population. However, it fails to incorporate population heterogeneities and individual interactions that may greatly affect the virus's rate of transmission.

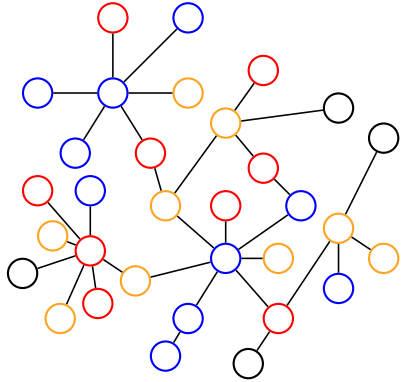
\begin{figure}
\begin{center}
\begin{tikzpicture}[-, >=stealth', auto, semithick, node distance=1.5cm]
\tikzstyle{every state}=[fill=white,draw=black,thick,text=black,scale=1,minimum height = 0.2cm, minimum width = 0.2cm]
\node[state, draw=red]  at (0,0)  (A)  {};
\node[state, draw=blue]   at (1,0) (B)   {};
\node[state, draw=blue]    at (0,-1) (C)  {};
\node[state, draw=jaunefonce]    at (1,-1) (D)  {};
\node[state, draw=blue]    at (-1,-1) (E)  {};
\node[state, draw=blue]    at (-0.5,-1.8) (F)  {};
\node[state, draw=red]    at (0.5,-1.8) (G)  {};
\node[state, draw=jaunefonce]    at (0.7,-2.5) (H)  {};
\node[state, draw=jaunefonce]    at (1.5,-1.4) (I)  {};
\node[state, draw=blue]    at (1.5,-3.2) (J)  {};
\node[state, draw=red]    at (1.5,-2.5) (JJ)  {};
\node[state, draw=blue]    at (1,-4) (K)  {};
\node[state, draw=jaunefonce]    at (0.3,-3.5) (L)  {};
\node[state, draw=red]    at (-0.3,-3.1) (M)  {};
\node[state, draw=blue]    at (-0.3,-2.3) (N)  {};
\node[state, draw=red]    at (-1,-2.3) (O)  {};
\node[state, draw=jaunefonce]    at (-0.8,-2.9) (P)  {};
\node[state, draw=black]    at (-1.2,-3.4) (Q)  {};
\node[state, draw=jaunefonce]    at (-0.7,-4) (R)  {};
\node[state, draw=red]    at (-0.2,-3.8) (S)  {};
\node[state, draw=blue]    at (2.5,-2.5) (T) {};
\node[state, draw=jaunefonce]    at (2.2,-3.2) (U) {};
\node[state, draw=red]    at (2.2,-4) (V) {};
\node[state, draw=black]    at (1.8,-4.6) (W) {};
\node[state, draw=blue]    at (0.7,-4.5) (X) {};
\node[state, draw=red]    at (2,-2) (Y) {};
\node[state, draw=red]    at (2,-0.7) (Z) {};
\node[state, draw=black]    at (3,-1.2) (ZZ) {};
\node[state, draw=black]    at (3.6,-1.6) (AA) {};
\node[state, draw=jaunefonce]    at (3,-2.8) (BB) {};
\node[state, draw=jaunefonce]    at (3.6,-3.2) (CC) {};
\node[state, draw=blue]    at (3,-3.6) (DD) {};
\path
(A)  edge[  left]       (C)
(B)   edge[ left]        (C)
(E)   edge[ left]        (C)
(F)   edge[ left]        (C)
(G)   edge[ left]        (C)
(I)   edge[ left]        (H)
(G)   edge[ left]        (H)
(J)   edge[ left]        (H)
(JJ)   edge[ left]        (J)
(L)   edge[ left]        (J)
(L)   edge[ left]        (M)
(N)   edge[ left]        (M)
(O)   edge[ left]        (M)
(P)   edge[ left]        (M)
(Q)   edge[ left]        (M)
(R)   edge[ left]        (M)
(S)   edge[ left]        (M)
(K)   edge[ left]        (J)
(U)   edge[ left]        (J)
(T)   edge[ left]        (J)
(V)   edge[ left]        (J)
(Z)   edge[ left]        (I)
(ZZ)   edge[ left]        (I)
(T)   edge[ left]        (Y)
(I)   edge[ left]        (Y)
(V)   edge[ left]        (W)
(BB)   edge[ left]        (CC)
(BB)   edge[ left]        (V)
(BB)   edge[ left]        (DD)
(BB)   edge[ left]        (AA)
(K)   edge[ left]        (X)
(C)  edge[ left]   (D);
\end{tikzpicture}
\caption[Network of PWID infected by HCV and their sharing partners]{\label{network}\textbf{Network of PWID infected by HCV and their sharing partners}.\\ Blue nodes represent susceptible individuals, yellow nodes represent individuals in the acute phase of HCV, red nodes represent individuals in the chronic phase of HCV and black nodes are individuals who died from HCV.}
\end{center}
\end{figure}

Regarding HCV, it has been stated that the primary route of transmission in the developed world is intravenous drug use (IDU)~\cite{maheshwari}. As people who inject drugs (PWID) tend to share drug injecting equipment with a limited number of partners (see for example the social network described in~\cite{rolls}), using a simple mass action model might by questionable.\\
Hence, assuming that individuals with the same number of sharing partners present the same risk in the infectious process, we opt for the heterogeneous mean field approach, first introduced in~\cite{net:hetero}. Each individual is assigned a degree (a number of partners) $k$, where $1\leqslant k \leqslant N-1$, and constitutes an element of the network (also referred to as a node, in graph theory). As individuals with $0$~partner do not contribute to the spread of the epidemic, only individuals with at least one partner will be considered. Figure~\ref{network} offers a visual representation of a static network of sharing partners infected by HCV. For every individual dying of HCV, it is assumed that a new susceptible individual enters the network, substituting for them.

Let us denote by $\rho$ the degree distribution, i.e. $\rho (k)$ is the proportion of individuals with degree $k$. Writing $s_k(t)$ (resp. $a_k(t)$ and $c_k(t)$) for the proportion of susceptible (resp. acute-HCV infected and chronic-HCV infected) individuals with degree $k$ at time~$t$, one has:
\begin{equation}
s(t)= \sum_{k=1}^{N-1} s_{k}(t), \qquad a(t)= \sum_{k=1}^{N-1} a_{k}(t), \qquad c(t)= \sum_{k=1}^{N-1} c_{k}(t).
\end{equation}
Referring now to \cite{net:hetero}, the proportion of infected partners for an individual with degree $k$ is given by:
\begin{eqnarray*}\theta_{k}(t) = \sum_{k'=1}^{N-1}\rho_N(k'|k) \left( a_{k'}(t) +c_{k'}(t)  \right) \end{eqnarray*}
where $\rho_N(k'|k)$ is the proportion of partners with degree $k'$ among all partners of an individual with degree $k$. Under the assumption that all nodes of a given degree are statistically equivalent, one finds that $
\rho_N(k'|k)=\left(k'\rho(k')\right)/\langle k \rangle$,
where $\langle k \rangle$ is the mean degree of the network~\cite{CompNet}. As $\rho_N(k'|k)$ then does not depend on $k$, we shall simply write $\rho_N(k')$. Thus (\ref{comparthcv}) becomes:
\begin{eqnarray}\label{comparthcv2}
\frac{  \mathrm{d}s_k(t)}{  \mathrm{d}t} &=& \Lambda- k \beta s_k(t) \theta_{k}(t) + p \sigma a_k(t) + \gamma c_k(t) -\mu s_k(t),  \\
\frac{  \mathrm{d}a_k(t)}{  \mathrm{d}t} &=& k \beta s_k(t) \theta_{k}(t) - (\sigma +\mu) a_k(t),\\
\frac{  \mathrm{d}c_k(t)}{  \mathrm{d}t} &=& (1-p) \sigma a_k(t) -(\gamma +\delta +\mu) c_k(t). 
\end{eqnarray}

To determine which degree distribution should be used to model the drug equipment sharing pattern between PWID, we refer to \cite{bobolski} and~\cite{pelude}. The authors study the social network of PWID in the Bushwick neighborhood in Brooklyn, New York, and analyse its degree distribution. This analysis showed that the resulting degree distribution could be assimilated to a power law distribution. Typical of scale-free networks, power law distributions reflect a preferential attachment dynamics \cite{netw:barabasi}, which is consistent with the fact that new PWID network members tend to join a particular injector network via someone who is already an influential part of it.

Henceforth, we assume that the social network of PWID admits a scale-free topology. In particular, one can write $\rho(k)\sim k^{-\alpha}$ where $\alpha$ is called scale parameter. However, according to \cite{clauset}, it appears that power laws with exponential cutoff tend to fit natural networks better than pure power laws, as almost surely there are no individuals with an extremely large number of partners. Thus, based on this result and for an arbitrary cutoff $\kappa$, one can write:\begin{eqnarray*}\rho(k)=\frac{k^{-\alpha} e^{-k/\kappa}}{\mathrm{Li}_{\alpha} \left( e^{-1/\kappa}\right)}  \end{eqnarray*} where $\mathrm{Li}_\alpha$ is the polylogarithm of order $\alpha$. 

\FloatBarrier

\subsection{Incorporating contact with HIV-infected individuals}\label{sidamodel}

The model developed in the previous subsection does not take into account the fact that among PWID, people infected by HCV are often co-infected by HIV (human immunodeficiency virus), which greatly increases the risk of death~\cite{coinf}.

Therefore, we introduce three new compartments representing individuals infected only by HIV ($S^*$), individuals in the acute phase of the HCV infection and infected by HIV ($A^*$) and individuals in the chronic phase of the HCV infection and infected by HIV ($C^*$). We now have $S(t)+A(t)+C(t)+S^*(t)+A^*(t)+C^*(t)=N$.

\begin{figure}
\begin{center}
\begin{tikzpicture}[->, >=stealth', auto, semithick, node distance=3.5cm]
\tikzstyle{every state}=[fill=white,draw=black,thick,text=black,scale=1,minimum height = 0.2cm, minimum width = 1.2cm]
\node[state]    (A)  {S};
\node[state]    (B)[right of=A]   {A};
\node[state]    (C)[right of=B]   {C};
\node[state]    (AA)  [below = 2cm of A]{S$^*$};
\node[state]    (BB)[right of=AA]   {A$^*$};
\node[state]    (CC)[right of=BB]   {C$^*$};
\coordinate[left = 1.0cm of A] (Z);
\draw[->] (Z)  to   node[above]   {$\Lambda$} (A);
\coordinate[below left = 0.4 and 0.4cm of A] (E);
\draw[->] (A)  to   node[right, pos=0.7]   {$\mu$} (E);
\coordinate[above right = 0.4 and 0.4cm of B] (F);
\draw[->] (B)  to   node[left, pos=0.7]   {$\mu$} (F);
\coordinate[above right = 0.4 and 0.4cm of C] (G);
\draw[->] (C)  to   node[left, pos=0.7]   {$\mu$} (G);
\coordinate[below left = 0.4 and 0.4cm of AA] (EE);
\draw[->] (AA)  to   node[right, pos=0.7]   {$\mu^*$} (EE);
\coordinate[below right = 0.4 and 0.4cm of BB] (FF);
\draw[->] (BB)  to   node[left, pos=0.7]   {$\mu^*$} (FF);
\coordinate[below right = 0.4 and 0.4cm of CC] (GG);
\draw[->] (CC)  to   node[left, pos=0.7]   {$\mu^*$} (GG);
\coordinate[right = 1.0cm of C] (DD1);
\draw[->] (C)  to   node[above]   {$\delta$} (DD1);
\coordinate[ right = 1.0cm of CC] (DD2);
\draw[->] (CC)  to   node[above]   {$\delta^*$} (DD2);
\path
(A)  edge[ bend left]   node[below]{$\beta$}    (B)
       edge[ left]   node[right=1em, pos=0.4]{$\beta^{\dagger}$}    (BB)
(B)   edge[ left]     node[below]{$\left(1-p\right)\sigma$}     (C)
        edge[ bend left]     node[above=0.3em]{$p\sigma$}     (A)
(C) edge[ bend right = 30]     node[above]{$\gamma$}     (A)
(AA)  edge[ bend left]   node[below]{$\beta$}    (BB)
(BB)   edge[ left]     node[above]{$\left(1-p^*\right)\sigma$}     (CC)
        edge[ bend left]     node[above]{$p^*\sigma$}     (AA)
(CC)   edge[ bend left = 30]     node[below, pos=0.2]{$\gamma^*$}     (AA)
(A)  edge[  left]   node[left]{$\beta^*$}    (AA)
(B)  edge[  left]   node[left]{$\beta^*$}    (BB)
(C)  edge[  left]   node[left]{$\beta^*$}    (CC);
\end{tikzpicture}
\caption[Compartmental model of HCV's natural history with HIV coinfection]{\label{markhiv}\textbf{Compartmental model of HCV's natural history with HIV coinfection}.\\ S, susceptible; A, acute HCV infection; C, chronic HCV infection. Starred compartments represent HIV infected individuals.}
\end{center}
\end{figure}
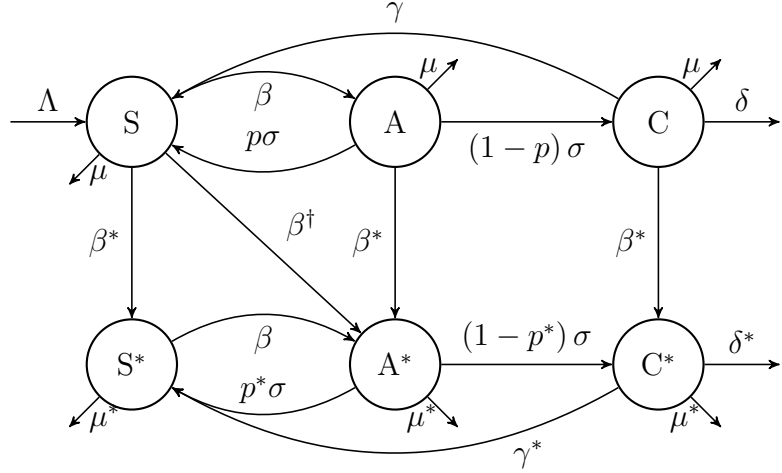

We also define $\gamma^*$ and $\delta^*$, the recovery and death rates of HCV for people already infected by HIV. As well as $\beta^*=\omega \times \pi^*$ the transmission rate of HIV, $\beta^{\dagger}=\omega \times \pi^{\dagger}$ the transmission rate of both HCV and HIV and $p^*$ the spontaneous clearance of HCV for people infected by HIV. Finally, we write $\mu^*$ for the mortality rate of PWID infected by HIV. A visual representation of the model is given in Figure~\ref{markhiv}. As previously, proportions of the different quantities are denoted with small letters.\\

We now introduce the probabilities that an individual with degree $k$ has an infected partner whichever the infection. Hence, we have the probability that an individual with degree $k$ has an HCV infected partner: \begin{eqnarray}\label{net:theta}\theta_{k}(t)= \displaystyle \sum_{k'=1}^{N-1}\rho_N(k'|k) \Big( a_{k'}(t) +c_{k'}(t) +a^*_{k'}(t) +c^*_{k'}(t)  \Big), \end{eqnarray} the probability that an individual with degree $k$ has an HIV infected partner:\begin{eqnarray}\label{net:thetastar}\theta^*_{k}(t) = \displaystyle \sum_{k'=1}^{N-1}\rho_N(k'|k) \Big(  s^*_{k'}(t) + a^*_{k'}(t) +c^*_{k'}(t)  \Big), \end{eqnarray} and the probability that an individual with degree $k$ has an HCV/HIV co-infected partner:\begin{eqnarray}\label{net:thetastarstar}\theta^{**}_{k}(t) = \displaystyle \sum_{k'=1}^{N-1}\rho_N(k'|k) \Big( a^*_{k'}(t) +c^*_{k'}(t)  \Big). \end{eqnarray} \\

From those equations and based on the model of Figure~\ref{markhiv}, one easily rewrites (\ref{comparthcv2}):
 \begin{eqnarray} \label{comparthiv}
\fl \frac{  \mathrm{d}s_k(t)}{  \mathrm{d}t} = \Lambda - k \beta s_k(t) \theta_k(t)  - k \beta^* s_k(t) \theta^*_k(t)  - k \beta^{\dagger} s_k(t)\theta^{**}_k(t)\nonumber\\ + p \sigma a_k(t) + \gamma c_k(t) - \mu s_k(t), \\
\fl \frac{  \mathrm{d}s_k^*(t)}{  \mathrm{d}t} = - k \beta s_k^*(t)  \theta_k(t) + k \beta^* s_k(t) \theta^*_k(t)  + p^*\sigma a_k^*(t) + \gamma^* c_k^*(t) - \mu^* s^*_k(t),\nonumber \\
\fl \frac{  \mathrm{d}a_k(t)}{  \mathrm{d}t} =  k\beta s_k(t)\theta_k(t) - (\sigma +\mu) a_k(t) -k \beta^*   a_k(t) \theta^*_k(t),\nonumber \\
\fl \frac{  \mathrm{d}a_k^*(t)}{  \mathrm{d}t} = k\beta s_k^*(t) \theta_k(t) + k \beta^{\dagger} s_k(t) \theta^{**}_k(t)   + k \beta^*  a_k(t) \theta^*_k(t)   - (\sigma + \mu^*)a_k^*(t),\nonumber \\
\fl \frac{  \mathrm{d}c_k(t)}{  \mathrm{d}t} = (1-p) \sigma a_k(t) - k \beta^* c_k(t) \theta^*_k(t) - (\gamma +\delta+\mu) c_k(t),\nonumber \\
\fl \frac{  \mathrm{d}c_k^*(t)}{  \mathrm{d}t} = (1-p^*)\sigma a_k^*(t) + k \beta^* c_k(t) \theta^*_k(t) - (\gamma^* +\delta^*+\mu^*) c_k^*(t).\nonumber
\end{eqnarray}

Such a framework allows for the study of an epidemic on a given network. However, when dealing with a large population, it might be wiser to deal with several subnetworks connected with each other. Indeed, when dealing with an epidemic, say at a country scale, computing a subnetwork for each city and allowing human movements between them might be more accurate than computing a single giant network. Based on this hypothesis, we propose to study, in the next section, the spread of the epidemic of HCV defined by (\ref{comparthiv}) across several subnetworks.

\FloatBarrier

\section{Epidemic propagation on a metapopulation model}

Metapopulation models, first described in \cite{levins}, consist in modeling the interactions of several spatially separated populations. Those models have been extensively used in the field of epidemiology to describe the spread of infectious diseases at a large scale by dividing the population into subpopulations corresponding to different households, cities, etc. Depending on human movements, a subpopulation, or patch, containing infected individuals is allowed to interact with other patches spreading the infection to neighboring subpopulations. We refer the reader for instance to \cite{meta2} and \cite{meta} for an illustration of the role of the global aviation network in the spread of the SARS infection.

Here we adapt this type of model to study the spread of HCV on a network of cities, assuming that the population of each city can be considered as a patch of a global metapopulation model. Each subpopulation admits a scale-free structure as defined in Section~\ref{hcvsection} and the spread of the virus is governed by the set of differential equations (\ref{comparthiv}). To keep things simple while making them concrete, we work in this section on networks of two and three cities.\\

\noindent\textbf{Two-city network.} Let us consider two cities $A$ and $B$ with injecting drug populations $N_A$ and $N_B$ respectively. Assuming that these populations are organized in scale-free networks, we denote respectively by $\langle k_A \rangle$ and $\langle k_B \rangle$ their mean degree and by $\rho_A(k)$ and $\rho_B(k)$ their degree distributions. We update the notations in (\ref{comparthiv}) by adding subscripts corresponding to cities, e.g. $s_{A}(t)$ and $s_{B}(t)$ denote the proportions of susceptible individuals in cities $A$ and $B$ respectively.

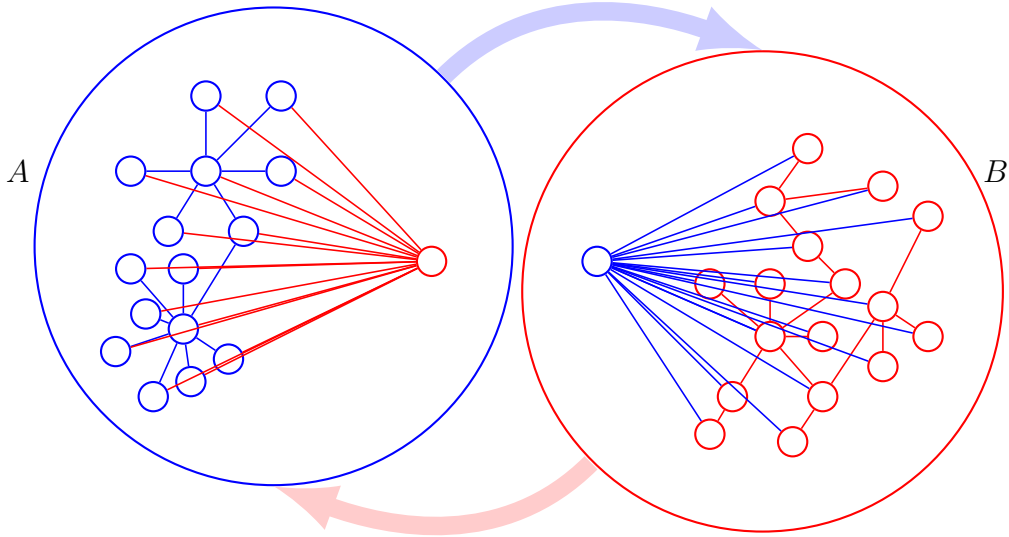
\begin{figure}
\begin{center}
\begin{tikzpicture}[-, >=stealth', auto, semithick, node distance=1.5cm]
\tikzstyle{every state}=[fill=white,draw=black,thick,text=black,scale=1,minimum height = 0.2cm, minimum width = 0.2cm]
\node[state, draw=blue]  at (-3,0)  (A)  {};
\node[state, draw=blue]   at (-2,0) (B)   {};
\node[state, draw=blue]    at (-3,-1) (C)  {};
\node[state, draw=blue]    at (-2,-1) (D)  {};
\node[state, draw=blue]    at (-4,-1) (E)  {};
\node[state, draw=blue]    at (-3.5,-1.8) (F)  {};
\node[state, draw=blue]    at (-2.5,-1.8) (G)  {};
\node[state, draw=red]    at (3.7,-2.5) (H)  {};
\node[state, draw=red]    at (4.5,-1.4) (I)  {};
\node[state, draw=red]    at (4.5,-3.2) (J)  {};
\node[state, draw=red]    at (4.5,-2.5) (JJ)  {};
\node[state, draw=red]    at (4,-4) (K)  {};
\node[state, draw=blue]    at (-2.7,-3.5) (L)  {};
\node[state, draw=blue]    at (-3.3,-3.1) (M)  {};
\node[state, draw=blue]    at (-3.3,-2.3) (N)  {};
\node[state, draw=blue]    at (-4,-2.3) (O)  {};
\node[state, draw=blue]    at (-3.8,-2.9) (P)  {};
\node[state, draw=blue]    at (-4.2,-3.4) (Q)  {};
\node[state, draw=blue]    at (-3.7,-4) (R)  {};
\node[state, draw=blue]    at (-3.2,-3.8) (S)  {};
\node[state, draw=red]    at (5.5,-2.5) (T) {};
\node[state, draw=red]    at (5.2,-3.2) (U) {};
\node[state, draw=red]    at (5.2,-4) (V) {};
\node[state, draw=red]    at (4.8,-4.6) (W) {};
\node[state, draw=red]    at (3.7,-4.5) (X) {};
\node[state, draw=red]    at (5,-2) (Y) {};
\node[state, draw=red]    at (5,-0.7) (Z) {};
\node[state, draw=red]    at (6,-1.2) (ZZ) {};
\node[state, draw=red]    at (6.6,-1.6) (AA) {};
\node[state, draw=red]    at (6,-2.8) (BB) {};
\node[state, draw=red]    at (6.6,-3.2) (CC) {};
\node[state, draw=red]    at (6,-3.6) (DD) {};
\node[state, draw=red]    at (0,-2.2) (AAA) {};
\node[state, draw=blue]    at (2.2,-2.2) (BBB) {};
\node[state, draw=none]    at (-5.5,-1) (CCC) {$A$};
\node[state, draw=none]    at (7.5,-1) (DDD) {$B$};
\node (GA)[circle, draw=blue, thick, fit=(B) (R) (AAA) (Q), inner sep=0pt] {};
\node (GB) [circle, draw=red, thick, fit=(Z) (X) (AA) (BBB), inner sep=0pt] {};
\draw[->, >=latex, blue!20!white,  line width=7pt] (GA) to [out=45, in=160] (GB.north) {};
\draw[->, >=latex, red!20!white,  line width=7pt] (GB) to [out=225, in=-20]  (GA.south) {};
\path
(A)  edge[color=blue,  left]       (C)
(B)   edge[color=blue, left]        (C)
(E)   edge[color=blue,  left]        (C)
(F)   edge[ color=blue, left]        (C)
(G)   edge[color=blue,  left]        (C)
(J)   edge[color=red,  left]        (H)
(JJ)   edge[ color=red,  left]        (J)
(M)   edge[color=blue,  left]        (G)
(L)   edge[color=blue,  left]        (M)
(N)   edge[color=blue,  left]        (M)
(O)   edge[ color=blue, left]        (M)
(P)   edge[ color=blue, left]        (M)
(Q)   edge[color=blue,  left]        (M)
(R)   edge[ color=blue, left]        (M)
(S)   edge[color=blue,  left]        (M)
(K)   edge[color=red,  left]        (J)
(U)   edge[color=red, left]        (J)
(T)   edge[ color=red,left]        (J)
(V)   edge[color=red, left]        (J)
(Z)   edge[ color=red,left]        (I)
(ZZ)   edge[ color=red,left]        (I)
(T)   edge[color=red, left]        (Y)
(I)   edge[ color=red,left]        (Y)
(V)   edge[color=red, left]        (W)
(BB)   edge[color=red, left]        (CC)
(BB)   edge[color=red, left]        (V)
(BB)   edge[ color=red,left]        (DD)
(BB)   edge[color=red, left]        (AA)
(K)   edge[color=red, left]        (X)
(C)  edge[color=blue,  left]   (D)
(AAA)  edge[color=red, left]   (A)
edge[color=red, left]   (B)
edge[color=red, left]   (C)
edge[color=red, left]   (D)
edge[color=red, left]   (E)
edge[color=red, left]   (F)
edge[color=red, left]   (G)
edge[color=red, left]   (L)
edge[color=red, left]   (M)
edge[color=red, left]   (N)
edge[color=red, left]   (O)
edge[color=red, left]   (P)
edge[color=red, left]   (Q)
edge[color=red, left]   (R)
edge[color=red, left]   (S)
(BBB)  edge[color=blue, left]   (J)
edge[color=blue, left]   (J)
edge[color=blue, left]   (I)
edge[color=blue, left]   (H)
edge[color=blue, left]   (JJ)
edge[color=blue, left]   (K)
edge[color=blue, left]   (T)
edge[color=blue, left]   (U)
edge[color=blue, left]   (V)
edge[color=blue, left]   (W)
edge[color=blue, left]   (X)
edge[color=blue, left]   (Y)
edge[color=blue, left]   (Z)
edge[color=blue, left]   (ZZ)
edge[color=blue, left]   (AA)
edge[color=blue, left]   (BB)
edge[color=blue, left]   (CC)
edge[color=blue, left]   (DD);
\end{tikzpicture}
\caption[Networks of $A$ and $B$ with additional nodes representing people from $A$ in $B$ and vice versa]{\label{network+}\textbf{Networks of $A$ and $B$ with the additional nodes representing people from $A$ in $B$ and vice versa}.\\ Network of the city $A$ (resp.~$B$) in blue (resp.~red) with the additional node of $B$ (resp.~$A$) in red (resp. in blue) interacting with every node. The blue light arrow represents the population travelling from $A$ to $B$ while the red light arrow represents the population from $B$ to $A$.}
\end{center}
\end{figure}

To model the effect of $B$ on the spread of the disease in $A$, we propose to add a node in the network of $A$ which is interacting with every node of $A$. In the same way, we add an extra node into the network of $B$ to model the effect of $A$ on $B$. These extra nodes allow the population of $A$ (resp. $B$) to connect with the new individuals moving from $B$ to $A$ (resp. from $A$ to $B$). A visual representation of the process is given in Figure \ref{network+}.

The extra node in the network of $A$ contains the main characteristics of $B$, i.e. the proportion of infected individuals in $B$ and  $\tau_{B,A}$  the number of people travelling from $B$ to $A$. Similarly, the extra node in the network of $B$ contains the main characteristics of $A$.
Hence, by focusing on the first equation of (\ref{comparthiv}), one can write:\begin{eqnarray}\label{equa}
\fl \frac{  \mathrm{d}s_{k,A}(t)}{  \mathrm{d}t} = - k \beta s_{k,A}(t) \theta_{k,A}(t)  - k \beta^* s_{k,A}(t) \theta^*_{k,A}(t)   - k \beta^{\dagger} s_{k,A}(t)\theta^{**}_{k,A}(t) - \mu s_{k,A}(t)\nonumber \\
 -  s_{k,A}(t)     \left( \frac{  \tau_{B,A}}{n_A+\tau_{B,A}}\right) \cdot    \bigg[ \beta  \Big( a_{B}(t) +c_{B}(t) +a^*_{B}(t) +c^*_{B}(t)  \Big) \bigg. \\
+ \bigg. \beta^* \Big( s^*_{B}(t) +a^*_{B}(t) +c^*_{B}(t)  \Big)  +  \beta^{\dagger}   \Big(a^*_{B}(t) +c^*_{B}(t)  \Big)   \bigg]    +p \sigma a_{k,A}(t)  + \gamma c_{k,A}(t)\nonumber
\end{eqnarray} where $n_A$ is the population still alive in city A. In particular, $n_A$ is defined as: $$n_A= s_A(t)+a_A(t)+c_A(t)+s^*_A(t)+a^*_A(t)+c^*_A(t).$$

\begin{figure}
\begin{center}
\begin{tikzpicture}[-, >=stealth', auto, semithick, node distance=1.5cm]
\tikzstyle{every state}=[fill=white,draw=black,thick,text=black,scale=1,minimum height = 0.2cm, minimum width = 0.2cm]
\node[state, draw=blue]  at (-3,0)  (A)  {};
\node[state, draw=blue]   at (-2,0) (B)   {};
\node[state, draw=blue]    at (-3,-1) (C)  {};
\node[state, draw=blue]    at (-2,-1) (D)  {};
\node[state, draw=blue]    at (-4,-1) (E)  {};
\node[state, draw=blue]    at (-3.5,-1.8) (F)  {};
\node[state, draw=blue]    at (-2.5,-1.8) (G)  {};
\node[state, draw=blue]    at (-2.7,-3.5) (L)  {};
\node[state, draw=blue]    at (-3.3,-3.1) (M)  {};
\node[state, draw=blue]    at (-3.3,-2.3) (N)  {};
\node[state, draw=blue]    at (-4,-2.3) (O)  {};
\node[state, draw=blue]    at (-3.8,-2.9) (P)  {};
\node[state, draw=blue]    at (-4.2,-3.4) (Q)  {};
\node[state, draw=blue]    at (-3.7,-4) (R)  {};
\node[state, draw=blue]    at (-3.2,-3.8) (S)  {};
\node[state, draw=red]    at (0,-2.2) (AAA) {};
\node[state, draw=teal]    at (-6.2,-1.2) (BBB) {};
\node[state, draw=none]    at (-7.5,-1) (CCC) {$A$};
\node (nik) [circle, draw=blue, thick, fit=  (AAA) (BBB), inner sep=2pt] {};
\coordinate[right = 3cm of AAA] (GA);
\coordinate[ left = 3cm of BBB] (GB);
\draw[->, >=latex, red!20!white,  line width=7pt] (GA) to [out=120, in=0] (nik.15) {};
\draw[->, >=latex, teal!20!white,  line width=7pt] (GB) to [out=-60, in=180]  (nik.195) {};
\path
(A)  edge[color=blue,    left]       (C)
(B)   edge[ color=blue,  left]        (C)
(E)   edge[color=blue,   left]        (C)
(F)   edge[ color=blue,  left]        (C)
(G)   edge[color=blue,   left]        (C)
(M)   edge[color=blue,   left]        (G)
(L)   edge[ color=blue,  left]        (M)
(N)   edge[color=blue,   left]        (M)
(O)   edge[color=blue,   left]        (M)
(P)   edge[color=blue,   left]        (M)
(Q)   edge[ color=blue,  left]        (M)
(R)   edge[ color=blue,  left]        (M)
(S)   edge[color=blue,   left]        (M)
(C)  edge[ color=blue,  left]   (D)
(AAA)  edge[color=red, left]   (A)
edge[color=red, left]   (B)
edge[color=red, left]   (C)
edge[color=red, left]   (D)
edge[color=red, left]   (E)
edge[color=red, left]   (F)
edge[color=red, left]   (G)
edge[color=red, left]   (L)
edge[color=red, left]   (M)
edge[color=red, left]   (N)
edge[color=red, left]   (O)
edge[color=red, left]   (P)
edge[color=red, left]   (Q)
edge[color=red, left]   (R)
edge[color=red, left]   (S)
(BBB)  edge[color=teal, left]   (A)
edge[color=teal, left]   (B)
edge[color=teal, left]   (C)
edge[color=teal, left]   (D)
edge[color=teal, left]   (E)
edge[color=teal, left]   (F)
edge[color=teal, left]   (G)
edge[color=teal, left]   (L)
edge[color=teal, left]   (M)
edge[color=teal, left]   (N)
edge[color=teal, left]   (O)
edge[color=teal, left]   (P)
edge[color=teal, left]   (Q)
edge[color=teal, left]   (R)
edge[color=teal, left]   (S);
\end{tikzpicture}
\caption[Network of $A$ with the additional nodes corresponding to cities $B$ and $C$]{\label{network++}\textbf{Network of $A$ with the additional nodes corresponding to cities $B$ and~$C$}.\\ The network of city $A$ is represented in blue. The additional nodes corresponding to cities $B$ and~$C$ are represented in red and green respectively. The red light arrow represents the population travelling from $B$ to~$A$ while the green light arrow represents the population travelling from $C$ to~$A$.}
\end{center}
\end{figure}
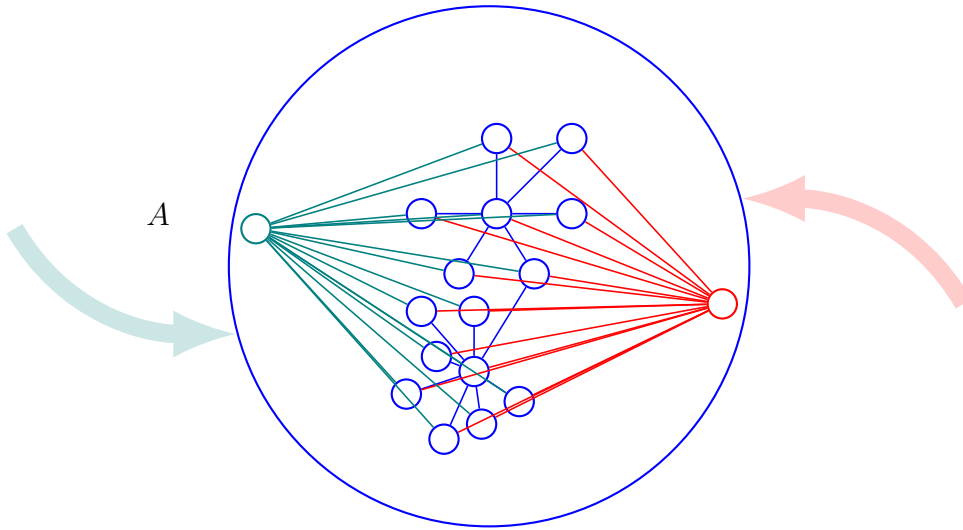

\noindent\textbf{Three-city network.} Let us introduce a third city $C$ with population $N_C$. As above, we assume that $C$ is organized in a scale-free network with mean degree $\langle k_C \rangle$ and degree distribution $\rho_C(k)$. To model the interactions between $A$ and $C$ (resp.~$B$ and~$C$), we add another node in the graph of $A$ (resp. $B$)  containing the main characteristics of $C$. See Figure \ref{network++} for a graphical visualization of the process in the particular case of~$A$.

Adding the new node corresponding to city $C$, equation (\ref{equa}) becomes:\begin{eqnarray}\label{net2:equa2}
\fl \frac{  \mathrm{d}s_{k,A}(t)}{  \mathrm{d}t} = - k \beta s_{k,A}(t) \theta_{k,A}(t)  - k \beta^* s_{k,A}(t) \theta^*_{k,A}(t)   - k \beta^{\dagger} s_{k,A}(t)\theta^{**}_{k,A}(t) - \mu s_{k,A}(t)\\
 \hspace{-0.5cm} - \displaystyle s_{k,A}(t) \sum_{\Phi \in\{B,C \}}    \left( \frac{  \tau_{\Phi,A}}{n_A+ \sum_{\Phi \in\{B,C \}}\tau_{\Phi,A}} \right) \cdot    \bigg[ \beta  \Big( a_{\Phi}(t) +c_{\Phi}(t) +a^*_{\Phi}(t) +c^*_{\Phi}(t)  \Big) \bigg. \nonumber \\
 + \bigg. \beta^* \Big( s^*_{\Phi}(t) +a^*_{\Phi}(t) +c^*_{\Phi}(t)  \Big)  +  \beta^{\dagger}   \Big(a^*_{\Phi}(t) +c^*_{\Phi}(t)  \Big)   \bigg]    +p \sigma a_{k,A}(t)  + \gamma c_{k,A}(t).\nonumber
\end{eqnarray} Full details of the complete set of differential equations are available in Appendix~\ref{ap:formula}.

\section{Simulations}

\subsection{Parameter values}\label{threecities}

Let us first consider three abstract cities $A$, $B$ and $C$ which PWID communities of 10000, 8000 and 5000 injecting drug users respectively. Cities with larger populations tend to be more attractive, so we use a constrained gravitational model~\cite{Gravity} to determine the proportion of a city's travelling population driven to a particular city among all other cities: \begin{eqnarray*} 
\tau_{A,B} =  \frac{ m_pn_A  n_B}{n_B+n_C}
\end{eqnarray*} where $m_p$ is the moving population percentage. Values of the different population movements were rounded up and reported in Table \ref{OD}. Those percentages were kept low to account for the fact that the population of a given city is far greater than the moving population.

  \begin{table*}[h!]
\centering
\ra{1.3}
\begin{tabular}{cccccccc}
\br
   & \multicolumn{7}{c}{Destination  }  \\
   \ns
& \crule{7} \\ 
\ns   & \multicolumn{3}{c}{$m_p = 0.1\%$}  &   &\multicolumn{3}{c}{$m_p = 1\%$}  \\
   \ns
   & \crule{3} & & \crule{3} \\
   \ns
  &  \multicolumn{1}{c}{$A$}&  \multicolumn{1}{c}{$B$}  &  \multicolumn{1}{c}{$C$} & & \multicolumn{1}{c}{$A$} &  \multicolumn{1}{c}{$B$} & \multicolumn{1}{c}{$C$} \\
$A$  & 0 &  6  & 4  &  &0 &  62  & 39   \\
$B$ & 5  & 0 &  3 &  & 53 &  0  & 27  \\
$C$ & 3 & 2 & 0 &  & 28 &  22  & 0   \\
\br
\end{tabular}
\caption[Annual origin-destination matrix]{\label{OD}\textbf{Annual origin-destination matrix.}}
\end{table*}

\begin{figure}[h!]
 \centering
\includegraphics{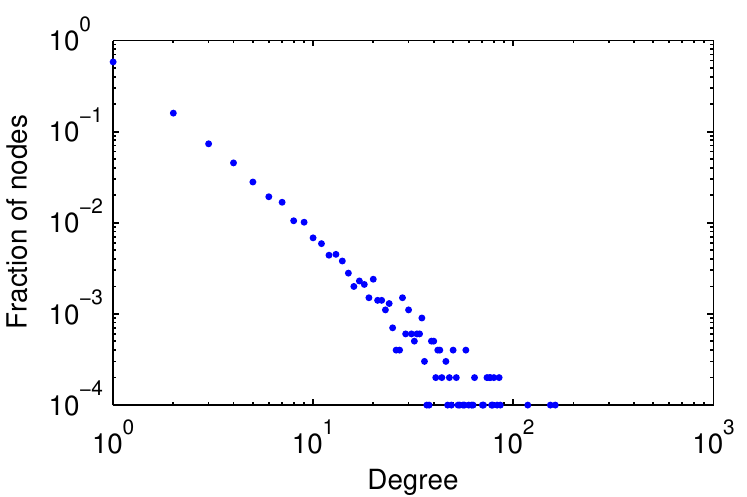}
\caption[Degree distribution of a scale-free network constituted of 10,000 nodes]{\label{powerlaw}\textbf{Degree distribution of a scale-free network constituted of 10,000 nodes.}\\  The degree distribution was computed for a network of 10,000 nodes  based on a power law distribution with exponential cutoff which parameters are $\alpha=1.8$ and $\kappa=40$.}
 \end{figure}

As far as network parameters are concerned, very limited data is available for their estimation. Hence, we use the values obtained by \cite{bobolski}, which were $\alpha=1.8$ for the scale parameter and $\langle k \rangle=3.0$ for the mean degree. Based on this, we initiate 1000 simulations of the degree distribution with $\alpha=1.8$ and different values of the exponential cut-off until reaching the value of 3.0 for the mean degree. The resulting cutoff was $\kappa=40$. Figure \ref{powerlaw} illustrates the degree distribution of a network of N=10,000.

As illustrated on Figure~\ref{powerlaw}, degrees may attain high values ($>10^2$). Since the model described in~\ref{ap:formula} is degree based, a high degree can lead to a very large amount of differential equations. Hence, for the sake of practicality, we decide to split the population into four arbitrary groups: people with degree 1, people whose degree lies between 2 and 5, people whose degree lies between 6 and 10 and people whose degree is  greater than 10.  \\

The values of the transmission risks of HCV and HIV are difficult to estimate. Several studies reported different values for the per contact probability of HCV transmission among PWID, ranging from 0.5\% to 10\% (see \cite{rolls, vicker, leary, boelen, bayoumi}). As for the transmission of HCV, we refer to \cite{rolls} who estimated a transmission risk of 1\% in a network of PWID in Australia. Considering the transmission of HIV, we refer to \cite{net:patel} who reported 63 transmissions for 10,000 exposures for the US epidemic. These transmission risks represent the probabilities of getting infected per act of sharing needles. Assuming that these probabilities are independent, the transmission risk $\pi^{\dagger}$  of both HCV and HIV is given by  $\pi^{\dagger}=\pi \times \pi^*$. 

As for the frequency of needle/syringe sharing, difficulties arise when one tries to obtain accurate estimations. Referring to \cite{vicker} we choose the value of 16 sharing acts per month. However, as this frequency is shared among the different partners of a given individual, we assume a per-partner sharing frequency of $16/3=5.33$, where 3 is the mean number of partners.

The values of the transmission risks introduced above reflect somehow the true HCV and HIV transmission risks for PWID. However, for illustrative purposes, we propose to temporary choose greater values for those risks by multiplying the transmission risks $\pi$, $\pi^*$ and $\pi^{\dagger}$ by some constant $\lambda$ to observe the impact of migrating people on the spread of the epidemic.

Considering the treatment of HCV, we decided to focus on the recent direct-acting antivirals (DAAs) which leads to high recovery rates. Indeed, based on a recent ANRS press release, the recovery rate of the new DAAs has been estimated up to $93\%$ for HCV/HIV co-infected patients (see \cite{anrs}). Similar results were observed for HCV mono-infected patients. Due to the fact that HCV is an asymptomatic disease, a large part of HCV infected patients are not aware of their serostatus. Globally, this involves a very low proportion of infected patients accessing to health care. Indeed, although France has one of the highest treatment rates in Europe, this proportion hardly reaches 5.2\% (see \cite{razavi}). 

As for the mortality rates of PWID infected by HCV, HIV or HCV and HIV, we refer to the work of \cite{santen} who calculated all-cause and cause-specific crude mortality rates (per 1000 person-years) and standardized mortality ratios for PWID in the Netherlands. A summary of the parameters used in the model are given in Table~\ref{rates}.

Since we want to quantify the impact of the interactions between cities, we assume that the epidemic starts in one of the three cities and observe the impact on the two others. We initiate the epidemic with a relatively low number of infected individuals in city $A$, that is $s_A(0)=0.985$, $c_A(0)=0.01$ (HCV mono-infected individuals in chronic phase) and $c^*_A(0)=0.005$ (HIV/HCV co-infected individuals in chronic phase). The values chosen for the initiation of the model are purely arbitrary. 

Due to the random nature of the network framework, we will focus on the average of each quantity over 1000 simulation runs.

  \begin{table*}[h!]
\centering
\caption[Parameters used in the model to simulate the spread of HCV]{\label{rates}\textbf{Parameters used in the model to simulate the spread of HCV}. }
\begin{tabular}{lcccc}
Parameter  & Notation & Value & Source \\
\textbf{Risk of transmission per contact} &  & &  \\
\quad HCV, \% & $\pi$  &  1  & \cite{rolls} \\
\quad HIV, \% & $\pi^*$ &  0.63 &   \cite{net:patel}  \\
\quad HCV and HIV, \% &$\pi^{\dagger}$ &    $6.3\times 10^{-5} $ & -- \\
\textbf{Semi-annual HCV spontaneous clearance} &  & &  \\
\quad HCV mono-infected, \% &$p$ &  26 & \cite{acute} \\
\quad  HCV/HIV co-infected, \% & $p^*$ &  10 & \cite{hernandez} \\
\textbf{Annual risk of death} &  & &  \\
\quad HCV/HIV uninfected, \textperthousand &$\mu$ &  10.4 &\cite{santen} \\
\quad HCV mono-infected, \textperthousand &$\delta$ & 22.7 & \cite{santen} \\
\quad HIV mono-infected, \textperthousand &$\mu^*$ & 44.3 & \cite{santen} \\
\quad  HCV/HIV co-infected, \textperthousand & $\delta^*$ &  54.9  & \cite{santen}  \\
\textbf{Network characteristics} &  & &  \\
\quad Mean degree & $\langle k \rangle$ & 3.0 & \cite{bobolski} \\
\quad  Scale parameter & $\alpha$ &  1.8  & \cite{bobolski}  \\
\quad  Exponential cutoff & $\kappa$ &  40  & --  \\
\textbf{Anti-HCV treatment characteristics} &  & &  \\
\quad Recovery rate, \%   &   $\nu$ & 93 & \cite{anrs}      \\
\quad Proportion of treated patients, \%    & $q$ & 5.2  &   \cite{razavi}  \\
\textbf{Other parameters} &  & &  \\
\quad Per-partner sharing frequency, per month  &$\omega$ & 5.33 & \cite{vicker} \\
\quad Duration of the acute phase, years  & $1/\sigma$ &   1/2  &  \cite{acute} \\
\quad Unit time, years  & -- &  1 & --  \\
\end{tabular}
\end{table*}

\subsection{Results on an abstract three-city network} 

\begin{figure}[h!]
     \begin{center}
    \begin{tikzpicture}
    \begin{customlegend}[legend columns=3,legend style={column sep=1ex},legend entries={\scriptsize{$m_p=0\% / \lambda=1$},\scriptsize{$m_p=0.1\% / \lambda=1$}, \scriptsize{$m_p=1\% / \lambda=1$}, \scriptsize{$m_p=0\% / \lambda=5$}, \scriptsize{$m_p=0.1\% / \lambda=5$},\scriptsize{$m_p=1\% / \lambda=5$},\scriptsize{$m_p=0\% / \lambda=10$},\scriptsize{$m_p=0.1\% / \lambda=10$},\scriptsize{$m_p=1\% / \lambda=10$}}]
    \addlegendimage{only marks, mark=o, draw=blue}
    \addlegendimage{only marks, mark=o, draw=red}
    \addlegendimage{only marks,mark=o, draw=darkvert}
    \addlegendimage{only marks,mark=x, draw=blue}
        \addlegendimage{only marks,mark=x,draw=red}
        \addlegendimage{only marks,mark=x, draw=darkvert}
    \addlegendimage{only marks,mark=+, draw=blue}
   \addlegendimage{only marks,mark=+, draw=red}
        \addlegendimage{only marks,mark=+, draw=darkvert}
    \end{customlegend}
\end{tikzpicture}
    \end{center}
    \vspace{-2mm}
    \centering
\includegraphics{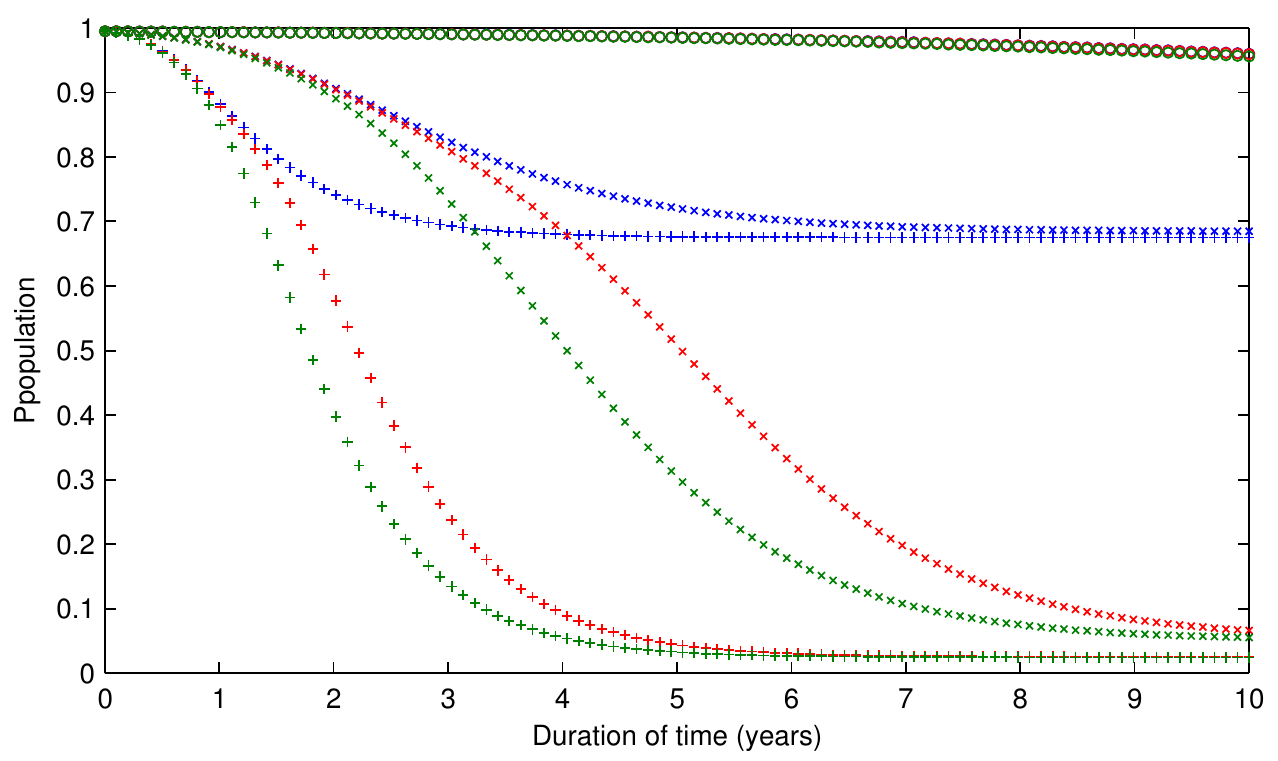}
\caption[Mean evolution of the susceptible population in the whole system]{\label{eqdiff}\textbf{Mean evolution of the susceptible population in the whole system.}\\  The different curves represent the mean quantities over $1000$ simulation runs. Blue curves represent the evolution of the susceptible population when no interactions between the cities are allowed ($m_p=0\%$),  red curves represent the evolution of the susceptible population assuming interactions up to 0.1\% of the populations ($m_p=0.1\%$) and green curves represent the evolution of the susceptible population assuming interactions up to 1\% of the populations ($m_p=1\%$).}
 \end{figure}

\begin{figure}[h!]
     \begin{center}
    \begin{tikzpicture}
    \begin{customlegend}[legend columns=3,legend style={column sep=1ex},legend entries={\scriptsize{$m_p=0\% / \lambda=1$},\scriptsize{$m_p=0.1\% / \lambda=1$}, \scriptsize{$m_p=1\% / \lambda=1$}, \scriptsize{$m_p=0\% / \lambda=5$}, \scriptsize{$m_p=0.1\% / \lambda=5$},\scriptsize{$m_p=1\% / \lambda=5$},\scriptsize{$m_p=0\% / \lambda=10$},\scriptsize{$m_p=0.1\% / \lambda=10$},\scriptsize{$m_p=1\% / \lambda=10$}}]
    \addlegendimage{only marks, mark=o, draw=blue}
    \addlegendimage{only marks, mark=o, draw=red}
    \addlegendimage{only marks,mark=o, draw=darkvert}
    \addlegendimage{only marks,mark=x, draw=blue}
        \addlegendimage{only marks,mark=x,draw=red}
        \addlegendimage{only marks,mark=x, draw=darkvert}
    \addlegendimage{only marks,mark=+, draw=blue}
   \addlegendimage{only marks,mark=+, draw=red}
        \addlegendimage{only marks,mark=+, draw=darkvert}
    \end{customlegend}
\end{tikzpicture}
    \end{center}
    \vspace{-2mm}
    \centering
\includegraphics{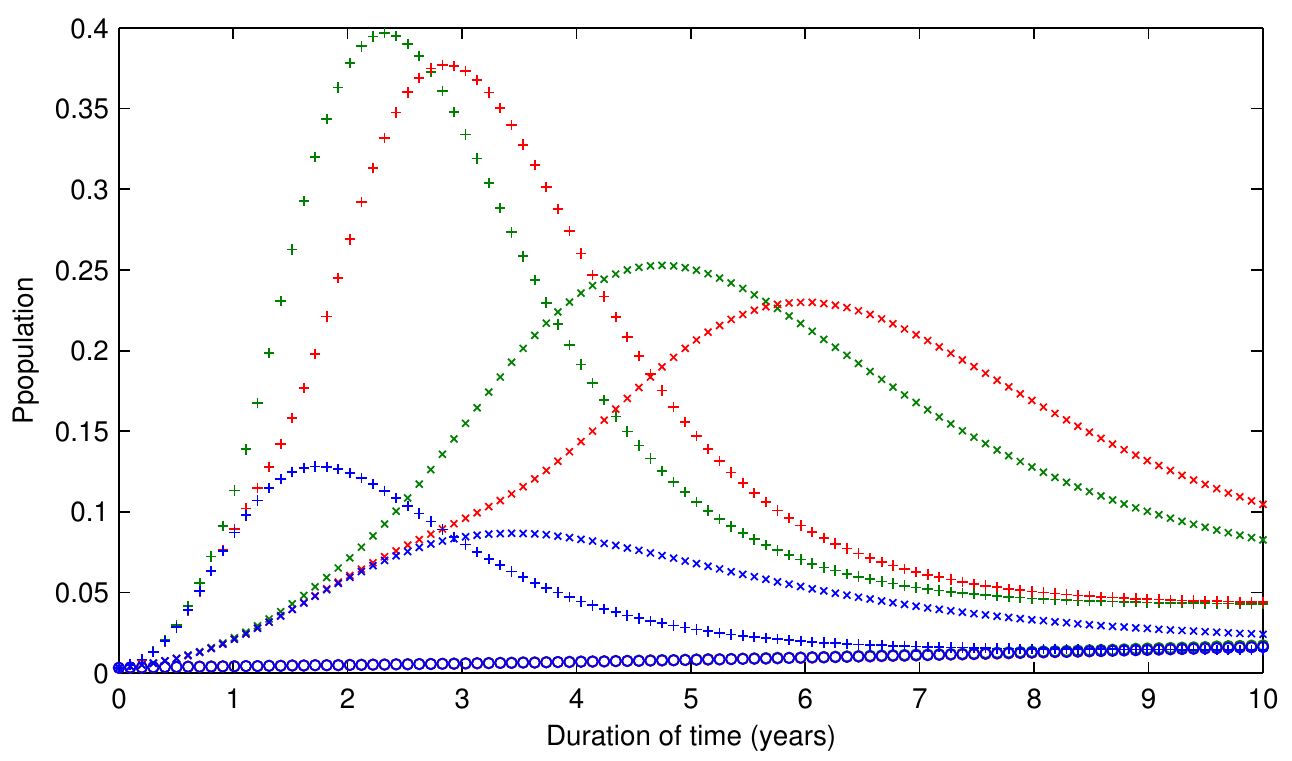}
\caption[Mean evolution of the HCV mono-infected population in the whole system]{\label{eqdiffmono}\textbf{Mean evolution of the HCV mono-infected population in the whole system.}\\  The different curves represent the mean quantities over $1000$ simulation runs. Blue curves represent the evolution of the HCV mono-infected population when no interactions between the cities are allowed ($m_p=0\%$),  red curves represent the evolution of the HCV mono-infected population assuming interactions up to $0.1$\% of the populations ($m_p=0.1\%$) and green curves represent the evolution of the HCV mono-infected population assuming interactions up to $1$\% of the populations ($m_p=1\%$).}
 \end{figure}

\begin{figure}[h!]
     \begin{center}
    \begin{tikzpicture}
    \begin{customlegend}[legend columns=3,legend style={column sep=1ex},legend entries={\scriptsize{$m_p=0\% / \lambda=1$},\scriptsize{$m_p=0.1\% / \lambda=1$}, \scriptsize{$m_p=1\% / \lambda=1$}, \scriptsize{$m_p=0\% / \lambda=5$}, \scriptsize{$m_p=0.1\% / \lambda=5$},\scriptsize{$m_p=1\% / \lambda=5$},\scriptsize{$m_p=0\% / \lambda=10$},\scriptsize{$m_p=0.1\% / \lambda=10$},\scriptsize{$m_p=1\% / \lambda=10$}}]
    \addlegendimage{only marks, mark=o, draw=blue}
    \addlegendimage{only marks, mark=o, draw=red}
    \addlegendimage{only marks,mark=o, draw=darkvert}
    \addlegendimage{only marks,mark=x, draw=blue}
        \addlegendimage{only marks,mark=x,draw=red}
        \addlegendimage{only marks,mark=x, draw=darkvert}
    \addlegendimage{only marks,mark=+, draw=blue}
   \addlegendimage{only marks,mark=+, draw=red}
        \addlegendimage{only marks,mark=+, draw=darkvert}
    \end{customlegend}
\end{tikzpicture}
    \end{center}
    \vspace{-2mm}
    \centering
\includegraphics{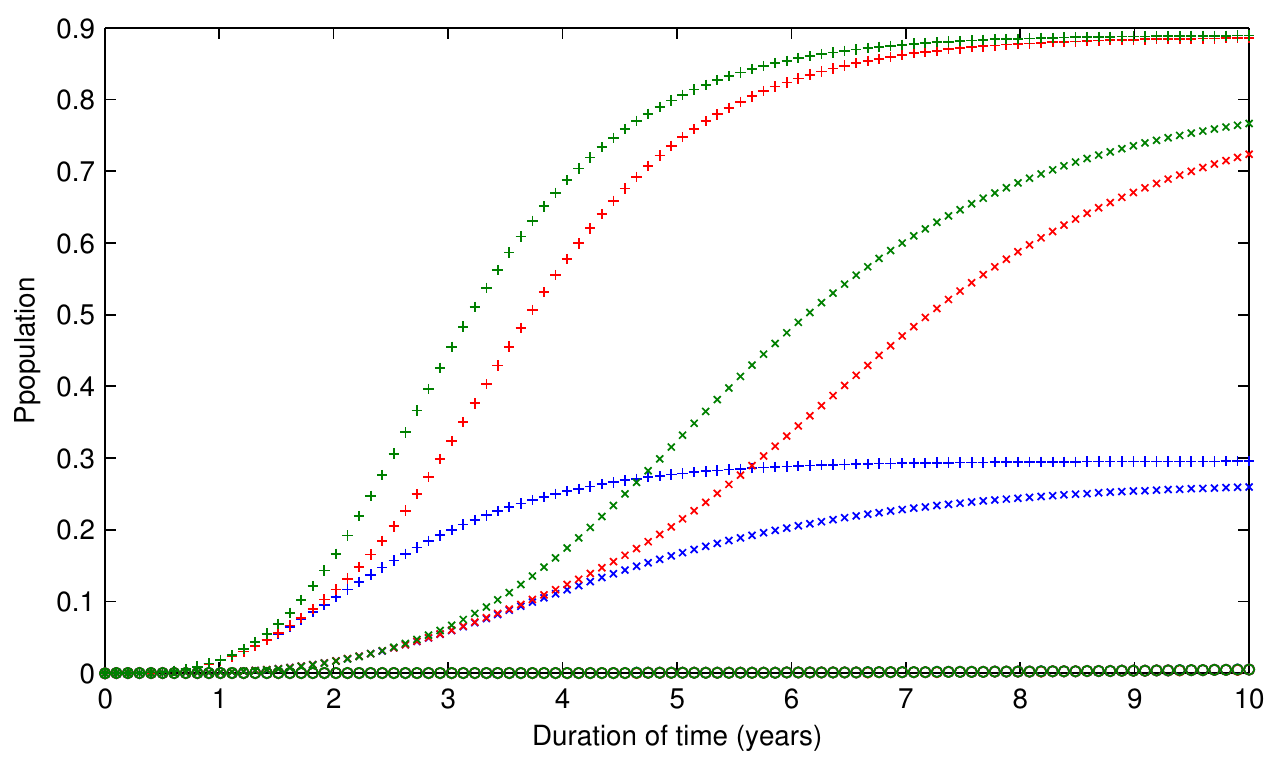}
\caption[Mean evolution of the HIV/HCV co-infected population in the whole system]{\label{eqdiffco}\textbf{Mean evolution of the HIV/HCV co-infected population in the whole system.}\\  The different curves represent the mean quantities over $1000$ simulation runs. Blue curves represent the evolution of the HIV/HCV co-infected population when no interactions between the cities are allowed ($m_p=0\%$), red curves represent the evolution of the HIV/HCV co-infected population assuming interactions up to $0.1$\% of the populations ($m_p=0.1\%$) and green curves represent the evolution of the HIV/HCV co-infected population assuming interactions up to $1$\% of the populations ($m_p=1\%$).}
 \end{figure}

Figures~\ref{eqdiff}, \ref{eqdiffmono} and \ref{eqdiffco} display the evolution of the susceptible, the HCV mono-infected and HIV/HCV co-infected populations in the whole system (cities $A$, $B$ and~$C$). The different curves of the three figures represent the population dynamics when no migration is assumed (blue markers), 0.1\% of the population is migrating (red markers) and 1\% of the population is migrating (green markers). Each quantity displayed in Figures~\ref{eqdiff}, \ref{eqdiffmono} and \ref{eqdiffco}  is the sum of the corresponding quantity over the four degree groups.

In Figure~\ref{eqdiff}, one sees that the susceptible population gradually depletes with the percentage of migrating population. As more and more people move from city~$A$, the epidemic spreads faster and has a stronger impact on cities $B$ and $C$. As expected, the spread of the epidemic is even stronger when coefficient $\alpha$ increases i.e. when the scale parameter of the degree distribution increases and individuals with larger degrees are present. One can see that the blue curves reach a minimum value between $60$\% and $70$\%. This phenomenon is due to the fact that the susceptible populations in cities $B$ and $C$ stay intact (no initial infected individual nor migration from infected cities).

In Figure~\ref{eqdiffmono}, one observes the same dynamics as in Figure~\ref{eqdiff}. The proportion of HCV mono-infected individuals in the population is greatly impacted by the percentage of migrating people and by the value of $\alpha$. Similarly, the blue curves here correspond to the proportion of HCV mono-infected individuals in the city $A$. This explains why that proportion is never higher than 33.3\%.

Lastly, Figure~\ref{eqdiffco} exhibits the same evolutions as the two previous figures. One can particularly observe on Figure~\ref{eqdiffco} that the evolution of the HIV/HCV co-infected population is greater than the evolution of the HCV mono-infected population of Figure~\ref{eqdiffmono}. This is due to the fact that, contrary to HCV, no treatment for HIV is considered in the model. Thus, each HCV mono-infected individual tends to get also infected by HIV as time goes by.\\

In the next section we study the spread of the infection on the network of France's $100$ largest urban centers. In particular, we examine the influence of the starting point of the epidemic.

\subsection{Results on the network of France's 100 largest urban areas}\label{popvhc}

We generalize the metapopulation model of Section~\ref{threecities} by studying the spread of HCV across a whole country: France. As we want our predictions to be relatively accurate without considering every single city, we choose to focus on the first hundred biggest urban areas in France (metropolitan France only) for a population of 43,187,838 individuals in 2012. Indeed, in \cite{rachlis} three main drivers of mobility for PWID were identified: legal problems, entering drug treatment programs, and drug tourism. We assume that the great majority of PWID tend to move towards and between the main urban areas. 
 
Then, the population of these urban areas is multiplied by the prevalence of PWID in France estimated at $5.9$ \textperthousand \, by the OFDT (\textit{Observatoire fran\c cais des drogues et des toxicomanies}, the French monitoring centre for drugs \& drug addiction) in 2006~\cite{ofdt}. Since this prevalence has been estimated for the population aged $15$ to $64$ ($63,8$\% of the total French population in 2012, i.e 40,433,870 individuals), we multiply the population of the French urban areas by $40433870/43187838=0.94$ for both populations to match. The original populations retained for the simulations are reported in Appendix~\ref{appendixpop}.\\
 
Regarding the origin-destination matrix, no data is available for the annual migration of PWID. Hence, we refer to a French database of residential mobility during a five-year period published on the website of INSEE (\textit{Institut national de la statistique et des \'etudes \'economiques}, the French office for national statistics).  Values are divided by 5 and rounded to obtain an annual origin-destination matrix. As we focus on PWID, values are multiplied by 5.9 \textperthousand.\\ 
 
To study the impact of migrating populations and of the size of the initial city infected on the spread of the infection, we shall consider two different starting points. We first choose Paris as a starting point since this city has the highest level of interactions with nearly all other French cities. To select the second starting point, we need a city which possesses a far lower level of interactions with all cities other than Paris. However, to prevent the snowball effect, it is better to also focus on a city which has very limited interactions with Paris. Hence, we choose Forbach, a small urban area in the North-Eastern part of France, as the starting point in a second set of simulations.

Referring to Section~\ref{threecities}, we set the proportion of susceptible individuals to $98.5$\% in the city where the epidemic starts.  The proportions of HCV mono-infected (chronic phase) and HIV mono-infected patients are set to $1$\% and  $0.5$\% respectively in that same city. All other cities are initiated at $100$\% of susceptible individuals.  Similarly, we assume that each urban area is organized according to a scale-free network, the degree distribution of which follows a power law with exponential cutoff and a multiplicative constant $\lambda$ of $10$. Parameters considered for this model are reported in Table \ref{rates}.\\

\begin{figure}[h!]
     \begin{center}
    \begin{tikzpicture}
    \begin{customlegend}[legend columns=4,legend style={column sep=1ex},legend entries={\scriptsize{Paris as starting point}, \scriptsize{Susceptible},\scriptsize{HCV mono-infected}, \scriptsize{HIV/HCV co-infected},\scriptsize{Forbach as starting point},  \scriptsize{Susceptible},\scriptsize{HCV mono-infected}, \scriptsize{HIV/HCV co-infected}}]
     \addlegendimage{only marks, mark=.}
    \addlegendimage{only marks, mark=+, draw=blue}
    \addlegendimage{only marks, mark=+, draw=red}
    \addlegendimage{only marks,mark=+, draw=darkvert}
     \addlegendimage{only marks, mark=.}
    \addlegendimage{only marks,mark=*, color=blue, draw=white}
        \addlegendimage{only marks,mark=*,color=red, draw=white}
        \addlegendimage{only marks,mark=*, color=darkvert, draw=white}
    \end{customlegend}
\end{tikzpicture}
    \end{center}
    \vspace{-2mm}
    \centering
\includegraphics{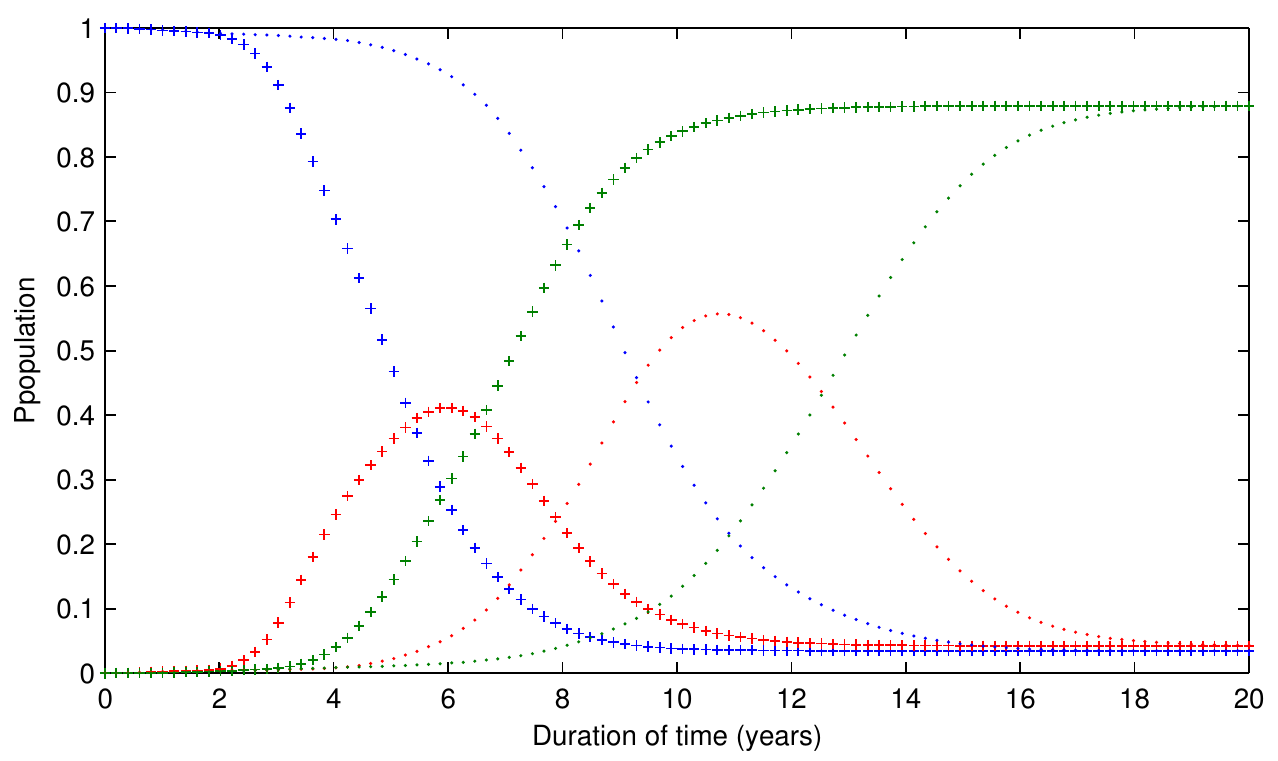}
\caption[Mean evolution of the population dynamics in France]{\label{eqdifftot}\textbf{Mean evolution of the population dynamics in France.}\\  The different curves represent the mean quantities over 1000 simulations. Dotted curves represent the evolution of the French population when Forbach is chosen as the epidemic starting point while (+) markers represent the evolution of the French population when Paris is chosen as the epidemic starting point. The blue markers represent the susceptible population while the red and the green markers represent the HCV mono-infected and the HIV/HCV co-infected populations respectively}
 \end{figure}

\noindent\textbf{Results.} Evolution of the different mean quantities are represented in Figure~\ref{eqdifftot} where the two different scenarios are exhibited in plain and dashed curves. One notices that, as expected, the epidemic spreads much more easily to the whole country when starting from Paris rather than Forbach. Indeed, as Paris has many more interactions with other urban areas than Forbach, the epidemic spreads faster. Particularly, one sees that the proportion of susceptible individuals depletes rapidly between the third and the eighth year when the epidemic starts in Paris while this proportion depletes between the seventh and the fourteenth year when the epidemic starts in Forbach. Similarly, one oberves that the proportion of HCV mono-infected patients increases over a few years and then decreases due to the superinfection with HIV.
 
\begin{figure}[h!]
     \begin{center}
    \begin{tikzpicture}
    \begin{customlegend}[legend columns=2,legend style={column sep=1ex},legend entries={\scriptsize{No infected individual},\scriptsize{Less than $20\%$ of infected individuals}, \scriptsize{From $20\%$ to $40\%$ of infected individuals}, \scriptsize{From $40\%$ to $60\%$ of infected individuals}, \scriptsize{From $60\%$ to $80\%$ of infected individuals},\scriptsize{More than $80\%$ of infected individuals}}]
    \addlegendimage{mark=*, color=vert, draw=white}
    \addlegendimage{mark=*, color=jaune, draw=white}
    \addlegendimage{mark=*, color=orange, draw=white}
    \addlegendimage{mark=*, color=orangefonce, draw=white}
        \addlegendimage{mark=*, color=rouge, draw=white}
        \addlegendimage{mark=*, color=rougefonce, draw=white}
    \end{customlegend}
\end{tikzpicture}
    \end{center}
\noindent \makebox[\textwidth]{\subfloat[One year after epidemic start]{\includegraphics[scale=0.7]{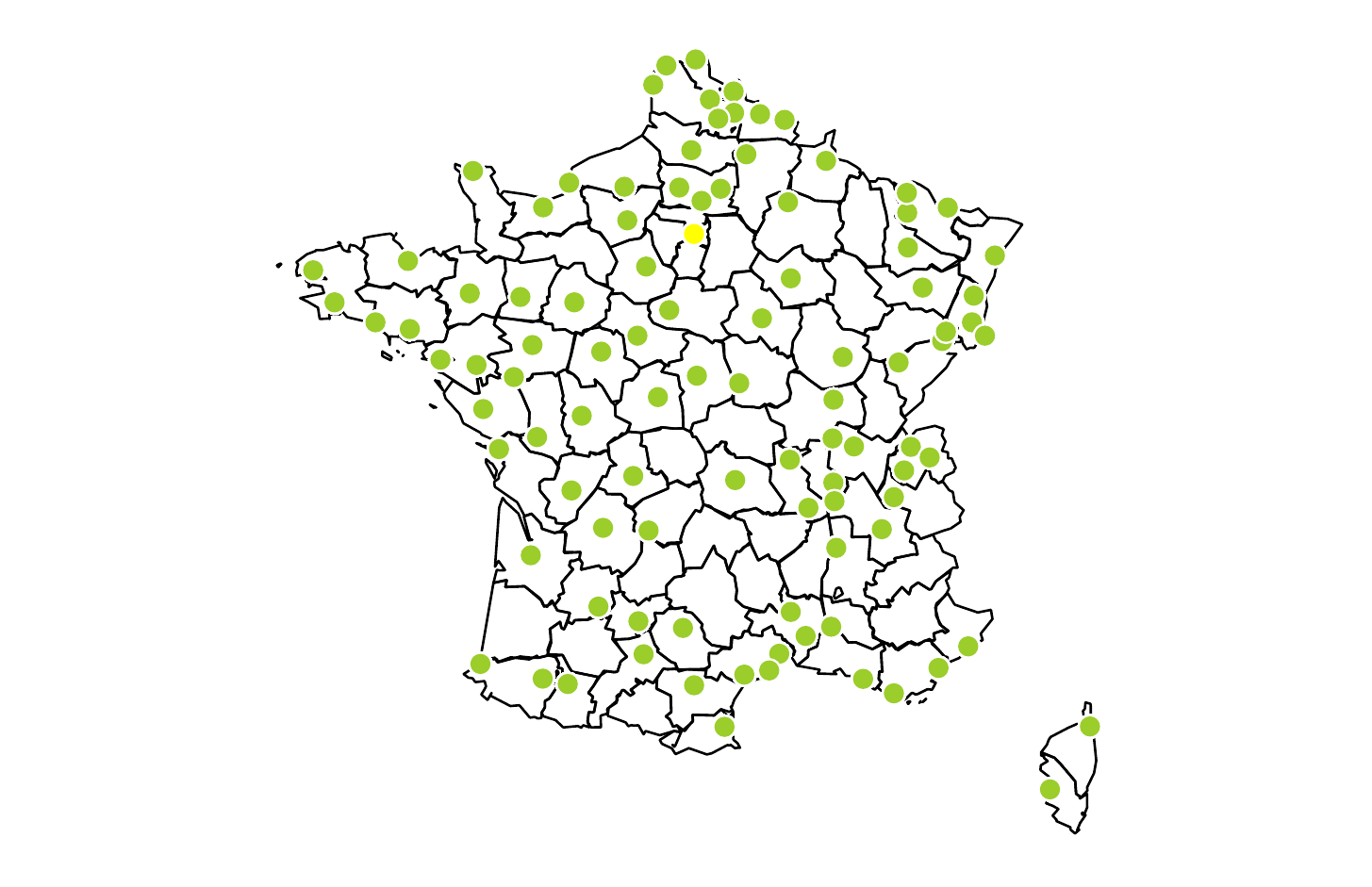}} \hspace{-2cm}\subfloat[Three years after epidemic start]{\includegraphics[scale=0.7]{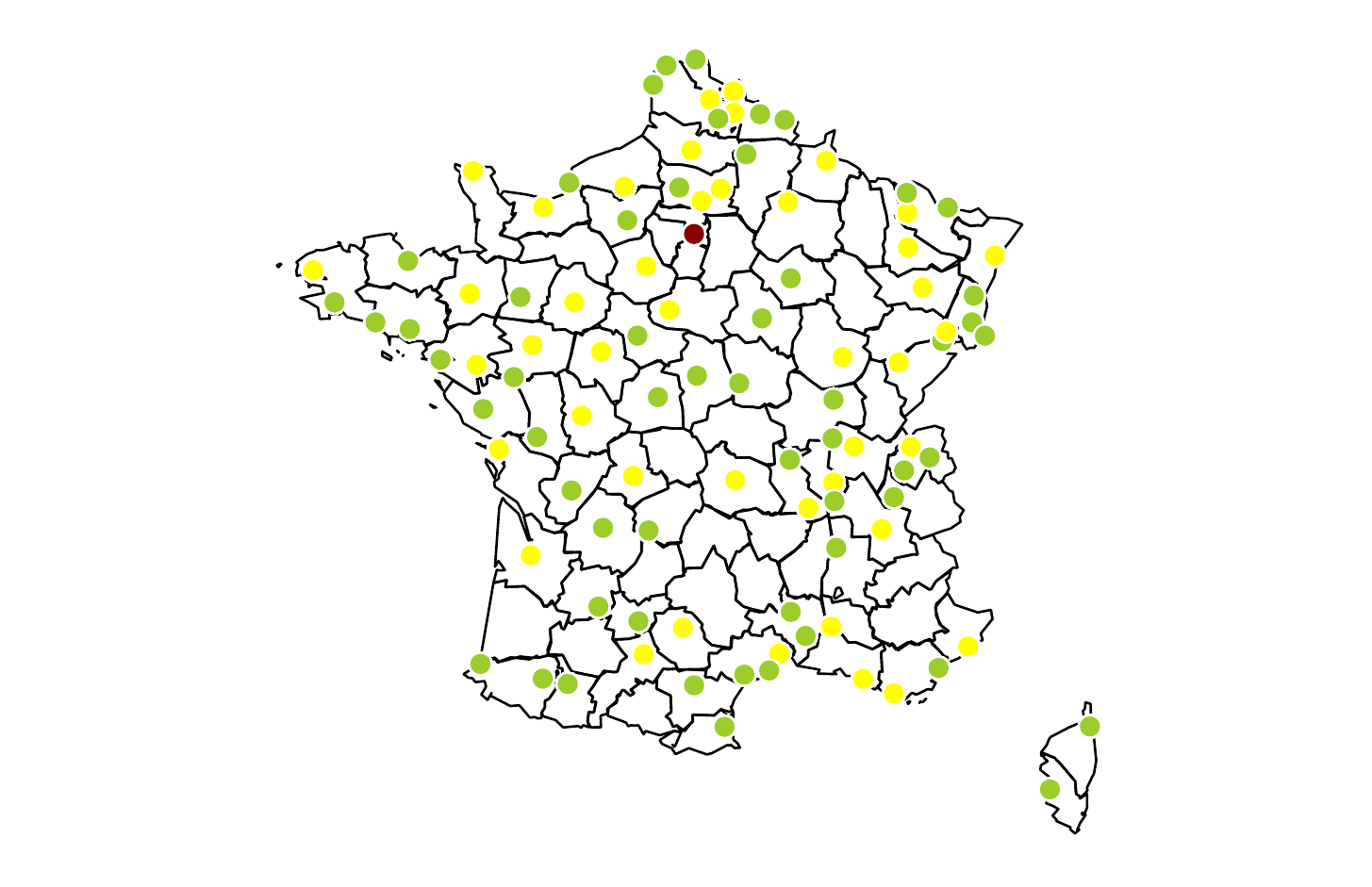}}}
\noindent \makebox[\textwidth]{\subfloat[Five years after epidemic start]{\includegraphics[scale=0.7]{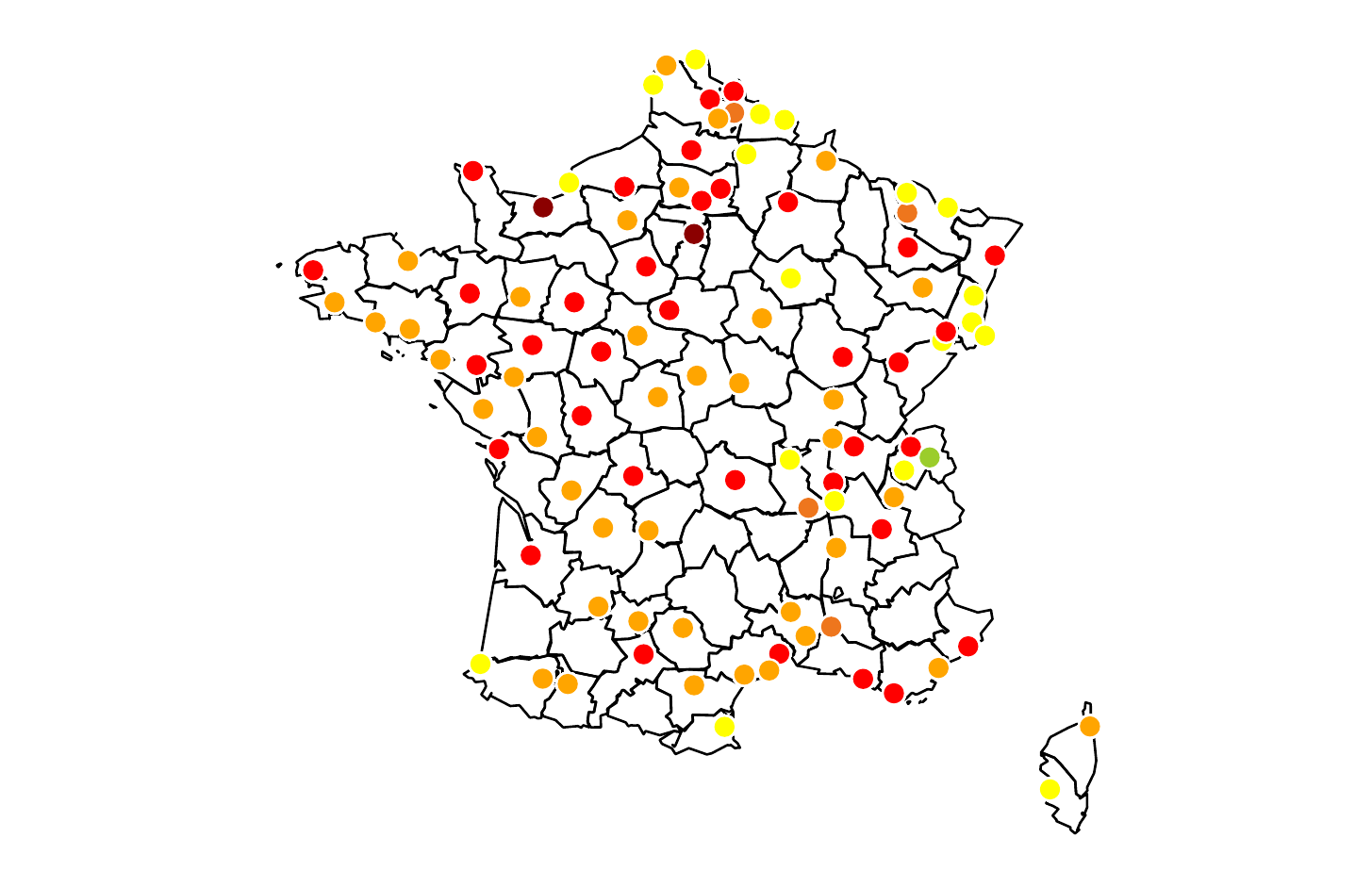}} \hspace{-2cm}\subfloat[Ten years after epidemic start]{\includegraphics[scale=0.7]{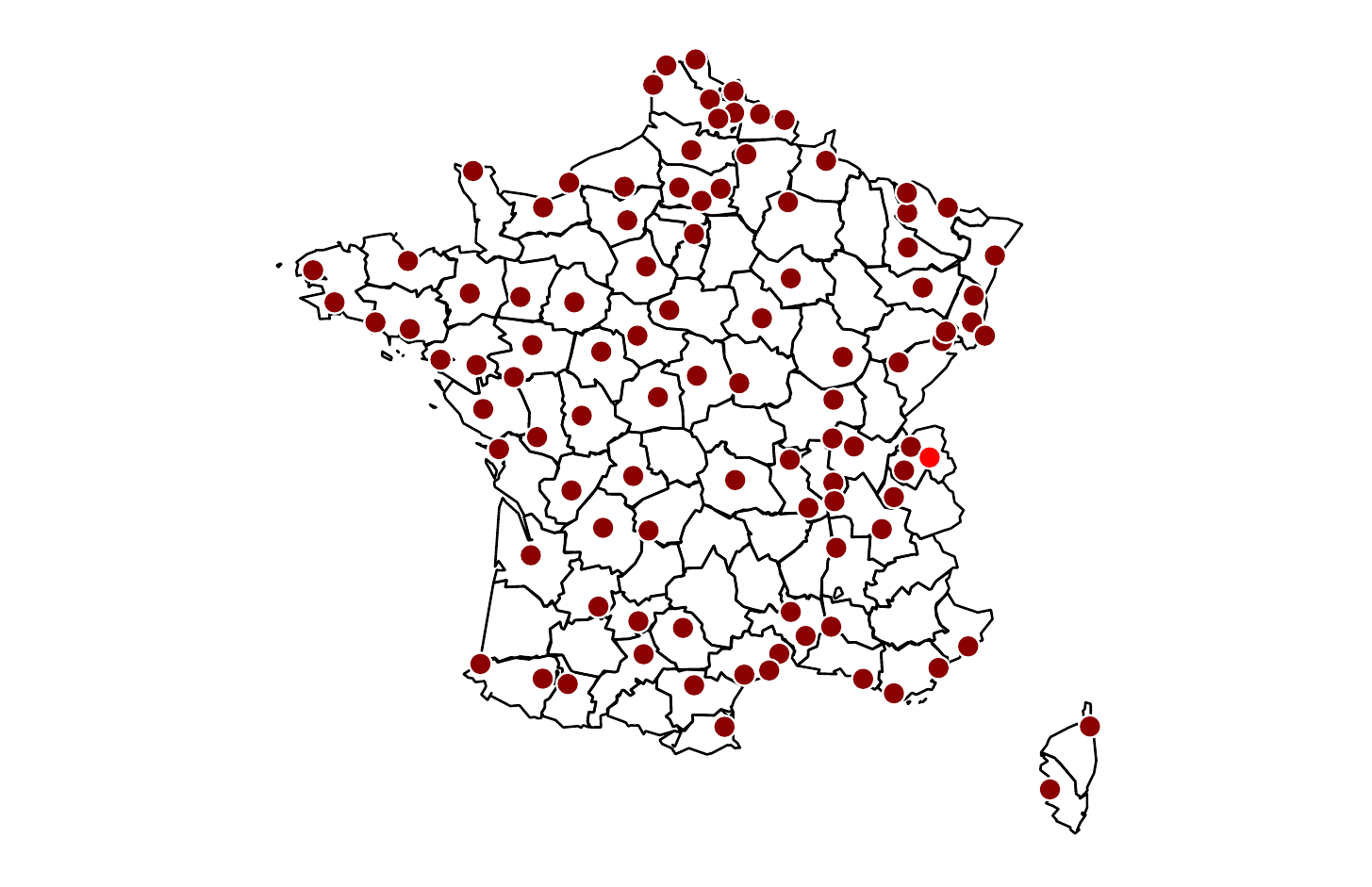}}}
\caption[Evolution of the spread of HCV in France with an epidemic start in Paris]{\label{paris}\textbf{Evolution of the spread of HCV in France with an epidemic start in Paris.}\\  Estimations of the proportions of HCV infected individuals in France over 1000 simulations. Each colored dot represents a French urban area. Starting point was 98.5\% of susceptible, 1\% of HCV infected individuals and 0.5\% of HIV infected individuals in Paris. Every other city started at 100\% of susceptible individuals.}
 \end{figure}

\begin{figure}[h!]
      \begin{center}
    \begin{tikzpicture}
    \begin{customlegend}[legend columns=2,legend style={column sep=1ex},legend entries={\scriptsize{No infected individual},\scriptsize{Less than $20\%$ of infected individuals}, \scriptsize{From $20\%$ to $40\%$ of infected individuals}, \scriptsize{From $40\%$ to $60\%$ of infected individuals}, \scriptsize{From $60\%$ to $80\%$ of infected individuals},\scriptsize{More than $80\%$ of infected individuals}}]
    \addlegendimage{mark=*, color=vert, draw=white}
    \addlegendimage{mark=*, color=jaune, draw=white}
    \addlegendimage{mark=*, color=orange, draw=white}
    \addlegendimage{mark=*, color=orangefonce, draw=white}
        \addlegendimage{mark=*, color=rouge, draw=white}
        \addlegendimage{mark=*, color=rougefonce, draw=white}
    \end{customlegend}
\end{tikzpicture}
    \end{center}
\noindent \makebox[\textwidth]{\subfloat[One year after epidemic start]{\includegraphics[scale=0.7]{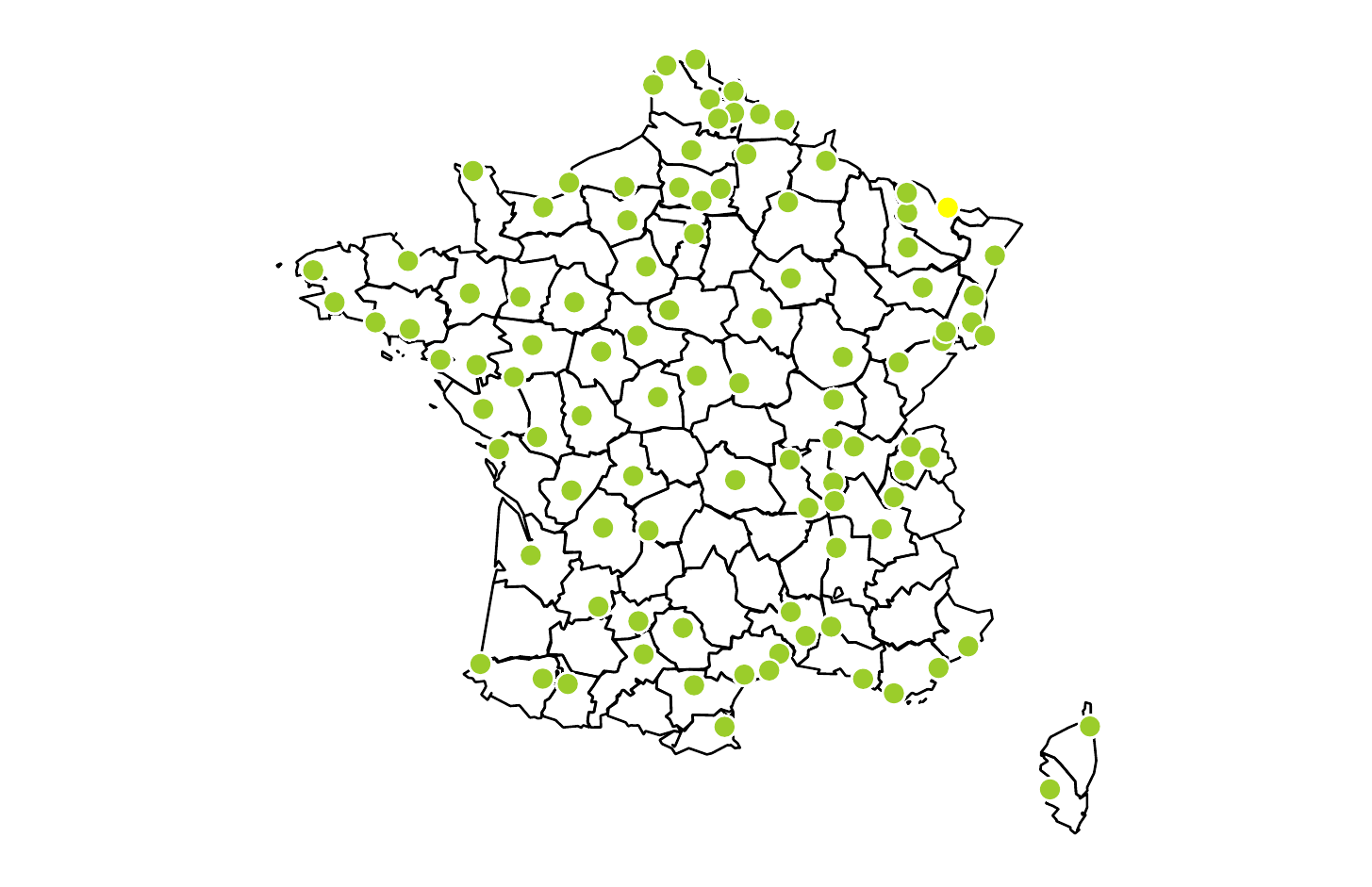}} \hspace{-2cm}\subfloat[Three years after epidemic start]{\includegraphics[scale=0.7]{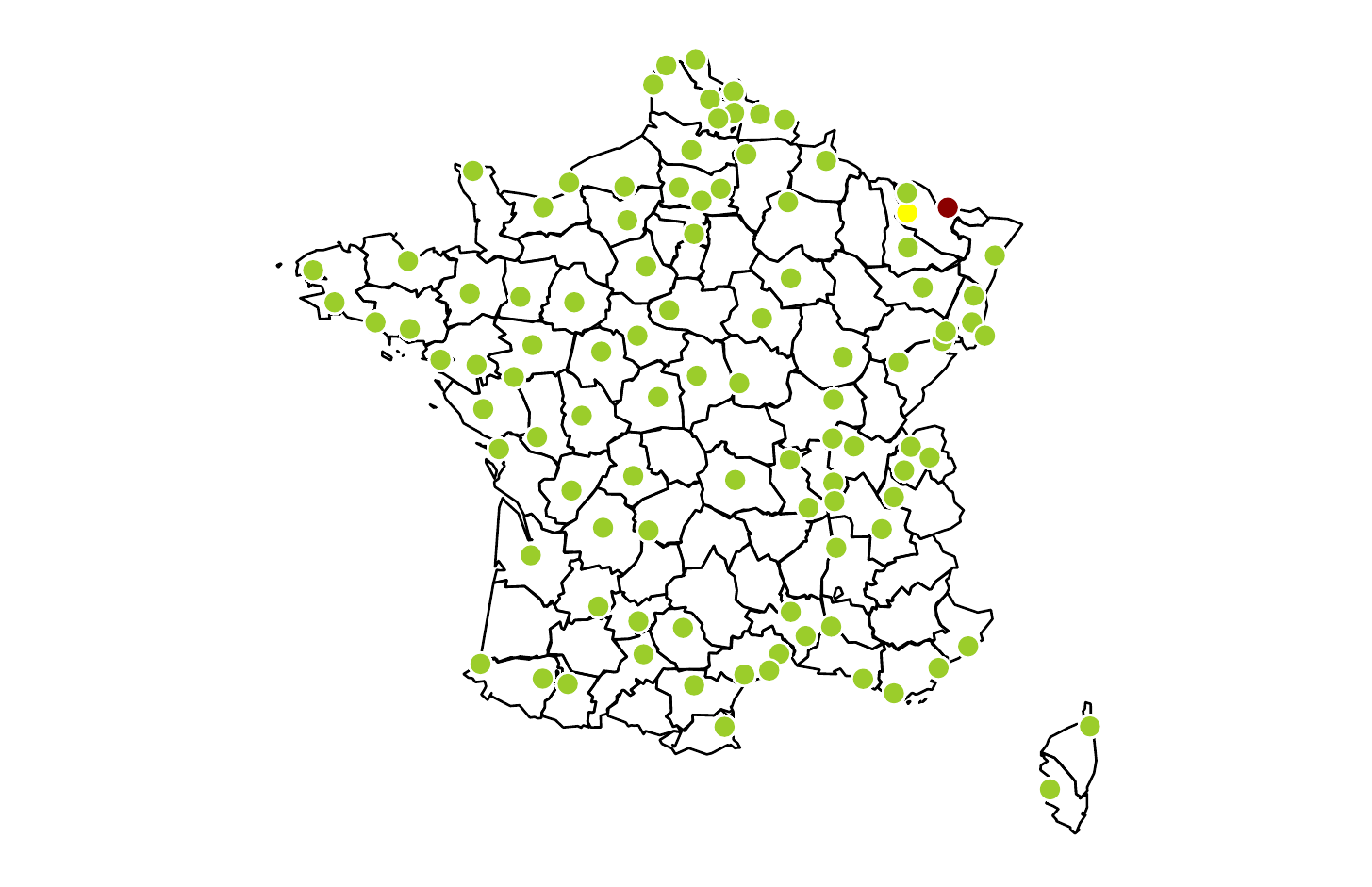}}}
\noindent \makebox[\textwidth]{\subfloat[Five years after epidemic start]{\includegraphics[scale=0.7]{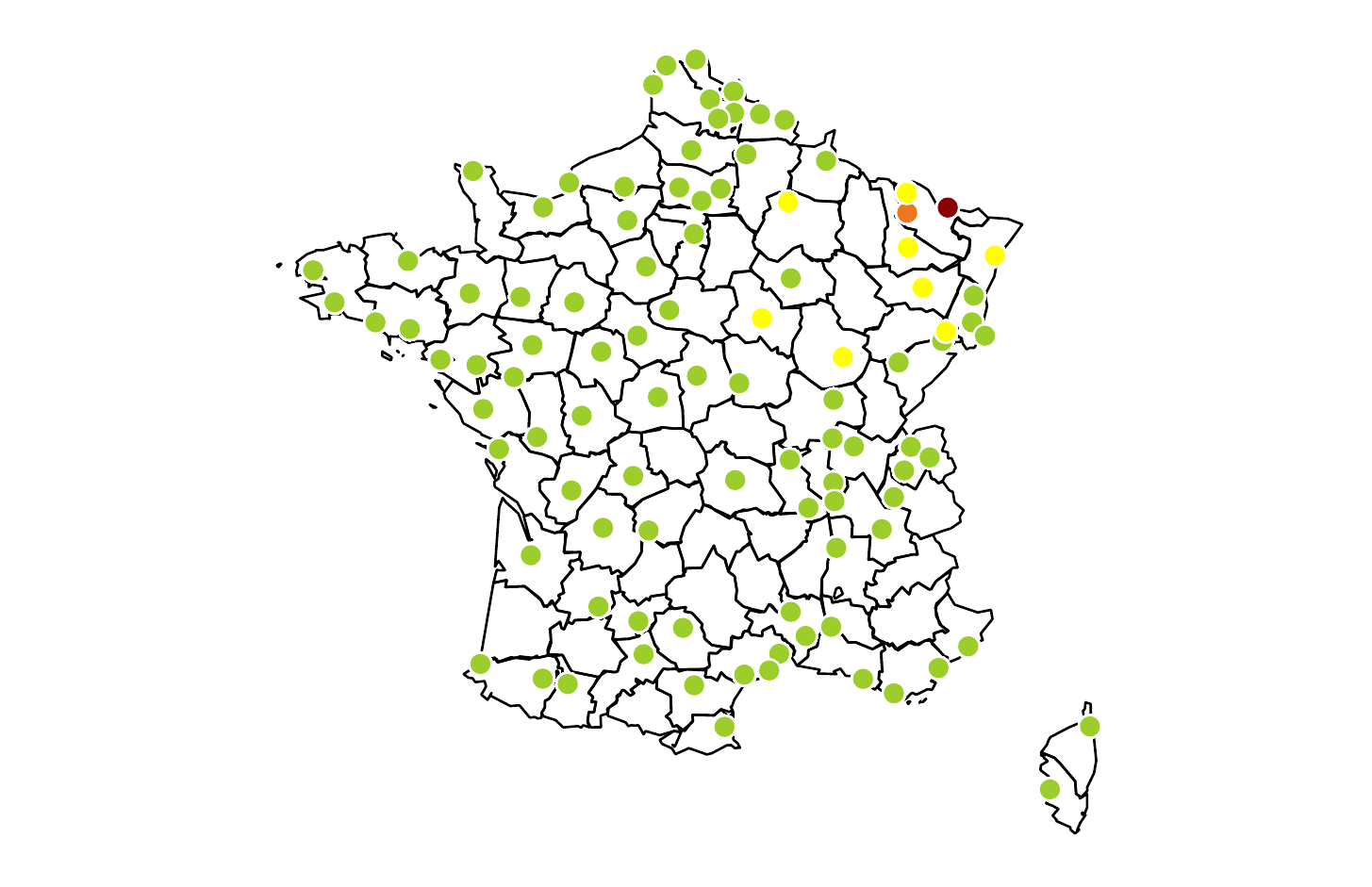}} \hspace{-2cm}\subfloat[Ten years after epidemic start]{\includegraphics[scale=0.7]{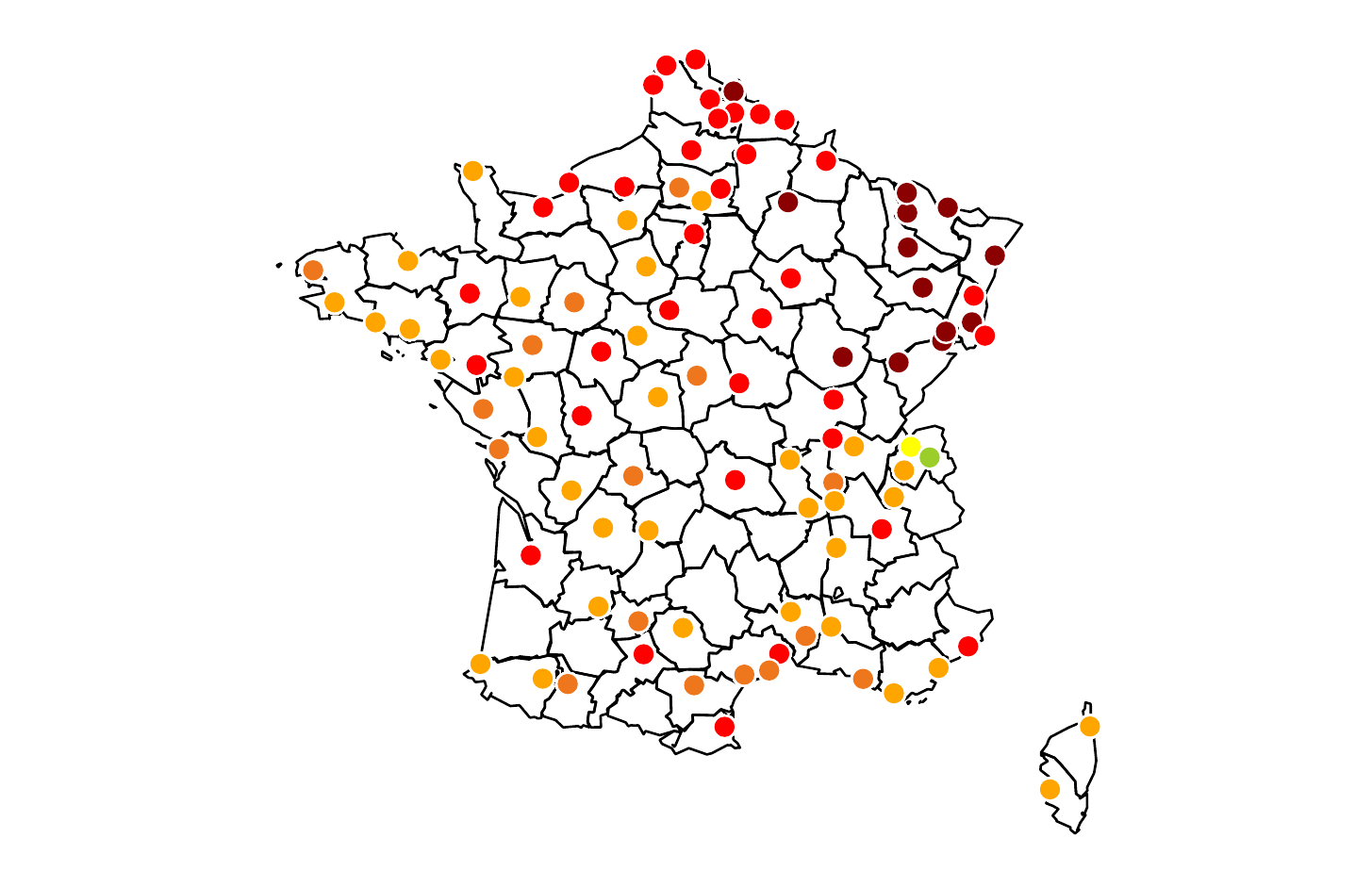}}}
\caption[Evolution of the spread of HCV in France with an epidemic start in Forbach]{\label{forbach}\textbf{Evolution of the spread of HCV in France with an epidemic start in Forbach.}\\  Estimations of the proportions of HCV infected individuals in France over 1000 simulations. Each colored dot represents a French urban area. Starting point was 98.5\% of susceptible, 1\% of HCV infected individuals and 0.5\% of HIV infected individuals in Forbach. Every other city started at 100\% of susceptible individuals.}
 \end{figure}

Figures \ref{paris} and \ref{forbach} give a comparison of the spread of the infection in the country. As intuitively expected, the epidemic spreads further and faster when starting in Paris. Indeed, ten years after the beginning of the epidemic, the high majority of urban areas admits a proportion of HCV infected individuals (mono-infection or co-infection) superior to $40$\% when the epidemic outbreak takes place in Paris. For comparison, when the epidemic breaks out in Forbach, not all urban areas are affected by the infection even after ten years. Finally, one notes that the last epidemic-affected city is Cluses in both cases. This can be explained by the fact that this urban area presents the lowest level of interaction with other French cities (it only interacts significantly with Gen\`eve-Annemasse).

\FloatBarrier

\section{Conclusion}
In this paper, we have introduced as a metapopulation model a compartmental model for the propagation of HCV on a network of networks, representing PWID populations in distinct but interconnected French cities.

Tested on a three-city network, it exhibits, as expected, an epidemic spread correlated with the amount of interactions between the cities: the epidemic grows faster and stronger when the amount of interactions between cities is high. Applied on a network of $100$ interconnected urban areas (corresponding to the $100$ lqrgest urban areas in France), the model yields very sensible stylized facts concerning the epidemic spread, in particular when chosing as starting points two different cities differing from each other by the amount of interactions they have with other urban areas. In particular, sensible differences are observed in terms of prevalence and propagation.

Our model combines two important features. First, it takes into account heterogeneities in the PWID population, which plays a crucial role in the transmission process. Secondly, it uses a metapopulation model to describe the infectious dynamics at a country scale.\\
Indeed, the use of a metapopulation model allows to divide the general PWID population into several PWID subpopulations and thus to better handle population heterogeneities at the subpopulation level with a contact network. Also, the model is relatively easy to handle, thanks to the use of extra-nodes working, in each network coorsedpong to a city, as proxys for individuals coming from other cities.\\

Having obtained sensible outcomes with this model, our next steps will be to obtain reliable data in order to calibrate and test it further. This is particularly challenging when dealing with PWID population. Indeed, whether it be needle sharing behavior for the estimation of HCV transmission or the degree of each individual for the establishment of a reasonable contact network, the lack of data forced us to make strong assumptions. Moreover, no cessation of treatment was taken into account despite the fact that the rate of non-compliance among PWID may be significant. Regarding the population moving from city to city, our only source was a French database of residential mobility which may misestimate the actual migrating dynamics of the PWID population.

It will also be of particular interest to explore part of the phase space arising when the values chosen for certain crucial parameters (such as the transmission rate) are allowed to vary across ranges away from their current reported or assumed typical value. This may reveal some unexpected transitions and critical values in the phase space corresponding to some of the parameters.

\newpage

\ack
This study was funded by a CIFRE grant to A.N. from the Bristol-Myers Squibb company, subsidized by the ANRT (CIFRE convention number: 2012/0879).

\section*{References}
\bibliographystyle{unsrt}
\bibliography{Bibliography}

\newpage

\appendix
\section*{Appendix}
\setcounter{section}{1}

\subsection{Differential equations}\label{ap:formula}

\begin{eqnarray*}
\fl \frac{  \mathrm{d}s_{k,\cdot}(t)}{  \mathrm{d}t} = \Lambda - k \beta s_{k,\cdot}(t) \theta_{k,\cdot}(t)  - k \beta^* s_{k,\cdot}(t) \theta^*_{k,\cdot}(t)   - k \beta^{\dagger} s_{k,\cdot}(t)\theta^{**}_{k,\cdot}(t) - \mu s_{k,\cdot}(t)\nonumber \\
 - \displaystyle s_{k,\cdot}(t) \sum_{\Phi }    \left( \frac{  \tau_{\Phi,\cdot}}{n_\cdot+ \sum_{\Phi} \tau_{\Phi,\cdot}}\right) \cdot    \bigg[ \beta  \Big( a_{\Phi}(t) +c_{\Phi}(t) +a^*_{\Phi}(t) +c^*_{\Phi}(t)  \Big) \bigg. \\
 + \bigg. \beta^* \Big( s^*_{\Phi}(t) +a^*_{\Phi}(t) +c^*_{\Phi}(t)  \Big)  +  \beta^{\dagger}   \Big(a^*_{\Phi}(t) +c^*_{\Phi}(t)  \Big)   \bigg]    +p \sigma a_{k,\cdot}(t)  + \gamma c_{k,\cdot}(t),\\
\fl \frac{  \mathrm{d}s_{k,\cdot}^*(t)}{  \mathrm{d}t} = - k \beta s_{k,\cdot}^*(t)  \theta_{k,\cdot}(t) + k \beta^* s_{k,\cdot}(t) \theta^*_{k,\cdot}(t)  + p^* \sigma a_{k,\cdot}^*(t) + \gamma^* c_{k,\cdot}^*(t) - \mu^* s^*_{k,\cdot}(t) \\
 + \displaystyle  \sum_{\Phi}    \left( \frac{  \tau_{\Phi,\cdot}}{n_\cdot+ \sum_{\Phi} \tau_{\Phi,\cdot}} \right) \cdot \bigg[  \beta^* s_{k,\cdot}(t)   \Big( s^*_{\Phi}(t) +a^*_{\Phi}(t) +c^*_{\Phi}(t)  \Big)   \bigg. \\
  \bigg. -\beta s^*_{k,\cdot}(t) \Big( a_{\Phi}(t) +c_{\Phi}(t) +a^*_{\Phi}(t) +c^*_{\Phi}(t)  \Big) \bigg],\\
\fl \frac{  \mathrm{d}a_{k,\cdot}(t)}{  \mathrm{d}t} =  k \beta s_{k,\cdot}(t) \theta_{k,\cdot}(t)   - (\sigma + \mu) a_{k,\cdot}(t) -k \beta^* a_{k,\cdot}(t) \theta^*_{k,\cdot}(t)\\
 +  \displaystyle \beta s_{k,\cdot}(t) \sum_{\Phi }    \left( \frac{  \tau_{\Phi,\cdot}}{n_\cdot+ \sum_{\Phi} \tau_{\Phi,\cdot}} \right) \cdot   \Big( a_{\Phi}(t) +c_{\Phi}(t) +a^*_{\Phi}(t) +c^*_{\Phi}(t)  \Big),\\
\fl \frac{  \mathrm{d}a_{k,\cdot}^*(t)}{  \mathrm{d}t} = k\beta s_{k,\cdot}^*(t) \theta_{k,\cdot}(t) + k \beta^{\dagger} s_{k,\cdot}(t) \theta^{**}_{k,\cdot}(t)   + k \beta^*  a_{k,\cdot}(t) \theta^*_{k,\cdot}(t)   -(\sigma + \mu^*) a_{k,\cdot}^*(t)\\
 +  \displaystyle \sum_{\Phi}    \left( \frac{  \tau_{\Phi,\cdot}}{n_\cdot+\sum_{\Phi} \tau_{\Phi,\cdot}} \right) \cdot  \bigg[  \beta s^*_{k,\cdot}(t)  \Big( a_{\Phi}(t) +c_{\Phi}(t) +a^*_{\Phi}(t) +c^*_{\Phi}(t)  \Big)  \bigg.\\
 +  \bigg.  \beta^* a_{k,\cdot}(t)    \Big(s^*_{\Phi}(t) + a^*_{\Phi}(t) +c^*_{\Phi}(t)  \Big) + \beta^{\dagger} s_{k,\cdot}(t)    \Big(a^*_{\Phi}(t) +c^*_{\Phi}(t)  \Big) \bigg],\\
\fl \frac{  \mathrm{d}c_{k,\cdot}(t)}{  \mathrm{d}t} = (1-p) \sigma a_{k,\cdot}(t) - k \beta^* c_{k,\cdot}(t) \theta^*_{k,\cdot}(t) - (\gamma +\delta +\mu) c_{k,\cdot}(t) \\
 - \displaystyle \beta^* ( a_{k,\cdot}(t)+ c_{k,\cdot}(t)) \sum_{\Phi }   \left( \frac{  \tau_{\Phi,\cdot}}{n_\cdot+\sum_{\Phi }\tau_{\Phi,\cdot}} \right) \cdot    \Big(s^*_{\Phi}(t) + a^*_{\Phi}(t) +c^*_{\Phi}(t)  \Big),\\
\fl \frac{  \mathrm{d}c_{k,\cdot}^*(t)}{  \mathrm{d}t} = (1-p^*)\sigma a_{k,\cdot}^*(t) + k \beta^* c_{k,\cdot}(t) \theta^*_{k,\cdot}(t) - (\gamma^* +\delta^* +\mu^*) c_{k,\cdot}^*(t) \\
 + \displaystyle \beta^* c_{k,\cdot}(t) \sum_{\Phi }   \left( \frac{  \tau_{\Phi,\cdot}}{n_\cdot+\sum_{\Phi }\tau_{\Phi,\cdot}} \right) \cdot    \Big(s^*_{\Phi}(t) + a^*_{\Phi}(t) +c^*_{\Phi}(t)  \Big).
\end{eqnarray*}
\newpage

\subsection{The main French urban areas}\label{appendixpop} 

\begin{table*}[h!]
\centering
\ra{1.3}
\tiny
\begin{tabular}{lcclc}
\br
 \multicolumn{1}{c}{Urban area}   & \multicolumn{1}{c}{Population} & & \multicolumn{1}{c}{Urban area}   & \multicolumn{1}{c}{Population}  \\
\mr
Paris  & 12,341,418 &  & Saint-Brieuc  & 171,721 \\
Lyon &    2,214,068 &  &  B\'eziers  &  165,498 \\
Marseille - Aix-en-Provence  &  1,727,070 & &  Montb\'eliard &  162,326 \\
Toulouse &    1,270,760  & & Niort & 152,721   \\
Lille$^{\dagger}$ &   1,166,452 &  & Vannes &   150,860 \\
Bordeaux  &   1,158,431 &  & Chartres & 145,735  \\
Nice &    1,004,914 &  & Bourges & 139,968   \\
Nantes &    897,713 &  &  Thionville  & 135,627  \\
Strasbourg$^{\dagger}$ &   768,868 &  &  Ch\^alon-sur-Sa\^one &  133,557 \\
Rennes &     690,467 &  & Boulogne-sur-Mer &    133,062 \\
Grenoble &   679,863  &  &  Maubeuge  & 129,931 \\
Rouen &   658,285 &    & Arras  &   128,784  \\
Toulon &   611,237  &  & Colmar   &  127,625 \\
Montpellier &   569,956   & & Blois  & 127,053 \\
Douai - Lens &    540,981 & &  Calais &   126,266 \\
Avignon &     515,536   &   & Quimper  & 125,487 \\
Saint-Etienne &   512,830   &  & Beauvais  &  125,095\\
Tours &   483,744   &  & Bourg-en-Bresse  &  122,806 \\
Clermont-Ferrand &     469,922 & &  Laval & 121,399   \\
Nancy &   434,479    &  &  La Roche-sur-Yon   &  117,965\\
Orl\'eans &   423,123  & &  Creil  &  117,654 \\
Angers  &    403,633  &  & Cherbourg-Octeville  & 116,517  \\
Caen  &   403,765  &  &  Tarbes    &   115,557  \\
Metz &   389,700  &   & Belfort  & 114,077 \\
Dijon &   377,590   & & Al\`es   & 113,769\\
B\'ethune &    368,633  & & Vienne &  112,334 \\
Valenciennes$^{\dagger}$ &     367,094  & & Agen  & 111,663  \\
Le Mans &   344,893 &   & Saint-Quentin &   111,474  \\
Reims &   317,611 & & Evreux    &   111,449 \\
Brest &   314,844  & & Roanne    &  107,209 \\
Perpignan &   309,962  & &  Charleville-M\'ezi\`eres     &  106,835 \\
Amiens  &   293,671   & &  Montauban    &  105,654 \\
Gen\`eve - Annemasse$^{\dagger}$  &   292,180   & & Cholet   & 104,917   \\
 Le Havre &    290,890   & & P\'erigueux   & 102,417 \\
 Bayonne$^{\dagger}$  &   288,359  &  &  Sarrebruck - Forbach$^{\dagger}$    & 101,806  \\
Mulhouse   &   284,739 &  &Nevers   &   101,586  \\
 Limoges  &    282,971  & &  Brive-la-Gaillarde   &101,435  \\
N\^imes &     259,348 & &   Ajaccio   &  100,643  \\
 Dunkerque &   257,773  & &   M\^acon   &  99,873  \\
Poitiers &    255,831    & &  Carcassonne & 97,801 \\
Besan\c con &   246,841 &  &     Albi  &   97,667  \\
Pau &   240,857  & & Compi\`egne  &97,502  \\
Annecy &   221,111  &  & Bastia    & 93,971 \\
 Chamb\'ery &   217,356 &  & Epinal   &   93,891  \\
   Lorient &    215,591 &  &  Fr\'ejus  &  93,562   \\
Saint-Nazaire &   213,083 & &   B\^ale - Saint-Louis$^{\dagger}$  & 93,018   \\
La Rochelle &    207,211  & &Châteauroux   &   92,723 \\
Troyes &     191,505 & & Auxerre  &   92,307 \\
Angoul\^eme &   180,593 &    & S\`ete &   91,101  \\
Valence &   175,636 & & Cluses  & 90,872  \\
\br
\end{tabular}
\captionof{table}[Population of the French urban areas based on census data for the year 2012]{\label{cities}\textbf{Population of the French urban areas based on census data for the year 2012}. \\
$^{\dagger}$ Only the French part is considered. }
\end{table*}

\end{document}